\documentclass[preprint,showpacs,preprintnumbers,amsmath,amssymb]{revtex4}

\usepackage{natbib}
\usepackage{graphicx}
\usepackage{dcolumn}
\usepackage{bm}

\begin{document}


\title{Non-Linear Gauge, Stochasticity and Confinement}

\author{Jose A. Magpantay}
\email{jose.magpantay11@gmail.com}
\affiliation{Quezon City 1101, Philippines\\}

\date{\today}

\begin{abstract}
I clarify, restate and show more clearly some key points I raised in a number of papers that discussed the non-linear gauge-fixing condition and quark confinement. I also correct some errors, which do not detract from the key findings, found in the original papers. However, there are two major corrections I will make in this paper, the first is on the proof of the Parisi-Sourlas mechanism and the second is on the effective action for the 'gluons', which leads to a direct proof of gluons being confined inside hadrons. The correction also leads to how the mass gap will be calculated, which was explicitly done in 2D. 

The starting point is that contrary to the prevailing ideas in the literature, the Coulomb gauge is an incomplete gauge-fixing condition in the sense that there are field configurations that cannot be gauge transformed to the Coulomb gauge. In other words the orbit of these configurations will not intersect the Coulomb gauge surface. I proposed the non-linear gauge condition precisely because it includes the Coulomb gauge in the high energy (short distance) regime and the quadratic regime (the large distance regime where the running coupling becomes large), where the gauge fields cannot be gauge transformed to the Coulomb surface. We proposed a new decomposition of the gauge potential in the non-linear regime, which involves an isoscalar (the divergence of the gauge field) and a new vector field, which exhibits a mass gap and confinement. When we add the quarks, we find that they are localized to a given distance scale and has an effective four-Fermi action with a linear potential. Thus, we have shown a mass gap for gluons and confinement for both dynamical quarks and gluons.           
\end{abstract}

\maketitle

\section{\label{sec:level1}Introduction}
Quantum chromodynamics (QCD) \cite{Fritsch} is well understood at the short distance regime \cite{Gross} \cite{Politzer}where perturbation expansion about the trivial vacuum makes sense because of the weak coupling. The calculations use a linear gauge-fixing, a Coulomb gauge for example, which clearly exposes the transverse gluons in the short distance regime. The running coupling, a perturbative computation, suggests that as the distance scale increases the coupling grows. This essentially limits the validity of the perturbation theory and non-perturbative physics must be used to explain the large distance regime of QCD, which shows the absence of colored states and the existence of a mass gap. There have been a number of ideas proposed to explain this regime and it is safe to say that there is no consensus yet as to which one of these explanations is correct.

In my work from the 1990s to 2004, I considered a different mechanism for confinement. The mechanism is based on a non-linear gauge that I proposed in the early 1990s. Although this idea of showing confinement via a choice of gauge seems to contradict the gauge-invariance of the result, this is not really a problem because all that the gauge choice does is to expose the physical degrees of freedom that are important in the appropriate regime. For example, the Coulomb gauge shows the transverse photons in electrodynamics in all distance scales and the transverse gluons appropriate in the short distance regime in QCD. Besides, the confinement mechanisms are based on some choice of gauge - abelian gauge for the monopoles, maximal abelian gauge for the vortices, Coulomb gauge for the Gribov copies, periodic gauge for calorons. It is a necessity that when we compute analytically in gauge theories, we have to specify a gauge. The same is true with lattice computations, which has been successful in computing for hadron masses to within a few percent of the actual values, see  \cite{McNeile} for a review, because propagators require a gauge choice. Thus, gauge fixing is inherent in computing in gauge theories. How does gauge-invariance follow from a computation that explicitly breaks the gauge symmetry? On the formal level, the Fadeev-Popov resolution of unity, where the gauge choice is compensated for by a corresponding Fadeev-Popov determinant, explicitly maintains gauge-invariance. At the practical level, as long as the different gauge choices somehow include the field configurations that give confinement, the gauge choices should manifest the same behavior for quantities like the Wilson loop, the Polyakov loop, mass gap and linear potential.     

My confinement mechanism begins with the observation that there are field configurations that cannot be gauge transformed to the Coulomb gauge. At the outset, I would like to state that this is contrary to the prevailing ideas in the literature \cite{DellAntonio}. This is discussed in Section II where we present the non-linear gauge. Consider field configurations that satisfy 
\begin{equation}\label{1}
(\partial \cdot D)(\partial \cdot A)=0.
\end{equation}
This gauge condition has two distinct regimes, (1) the short-distance where the running coupling goes to zero making $\partial \cdot D \rightarrow \partial^2$, a positive definite operator thus giving $\partial \cdot A=0$, which is the Coulomb gauge, and (2) the large distance regime, where the running coupling is strong, $ \partial \cdot D $ becomes singular with $ \partial \cdot A $ is its zero mode. In previous papers \cite{Magpantay1}, \cite{Magpantay2} I showed that the two regimes are non-intersecting, which means those that satisfy the Coulomb gauge cannot be gauge transformed to the non-linear regime and vice versa. This proves that the Coulomb gauge is insufficient because it will miss field configurations that may be significant in showing a physical effect. And indeed, this is what I showed in subsequent papers.

The first consequence of the non-linear gauge defined by equation (1) is it leads to a particular decomposition of the Yang-Mills potential as shown in \cite{Magpantay3}. The decomposition involves new degrees of freedom, which we discuss in Section III. These are an isospin set of scalars $ f^{a}(x)=\partial \cdot A^{a}(x) $ and a vector field $ t_{\mu}^{a}(x) $. Since we have more degrees of freedom, there is an extra constraint that these fields must satisfy. 

Section IV starts with a discussion on the significance of the transverse degrees of the gauge field. In the non-linear gauge, the gauge potentials are not transverse and the isoscalar fields $ \partial \cdot A^{a} = f^{a} \neq 0 $, will play a crucial role in the non-perturbative regime. In this section, we focus on the pure $ f^{a} $ dynamics. We find that all spherically symmetric fields (we consider Euclidean space-time, $ x = (x_{\mu}x_{\mu})^{\frac{1}{2}} $ are classical solutions and I proposed to treat these solutions as a stochastic field with a white noise distribution. 

Section V discusses the three hints of confinement that follow from the ideas in Section IV. The first is a heuristic derivation of a linear potential between quarks. The second is to show the area law behavior of the Wilson loop \cite{Magpantay3}, essentially showing again a linear potential for static quarks. The third is that if we consider the pure $ f^{a} $ dynamics, we find that it is essentially a 4D O(1,3)sigma model in a random field thus exhibiting a stochastic supersymmetry and by the Parisi-Sourlas mechanism is equivalent to a 2D O(1,3) sigma model \cite{Magpantay4}. Unfortunately, the proof of the Parisi-Sourlas mechanism in that paper has an error, which is corrected in Section V. The result is the same, the pure $ f^{a} $ dynamics is equivalent to a 4D O(1,3) sigma model in random magnetic field thus exhibiting a stochastic supersymmetry thus equivalent to an O(1,3) sigma model in 2D.  

But what is confined are dynamical quarks and gluons so we need to extend the hints of confinement - for static quarks (Wilson loop) and for isoscalar fields, the divergence of the gluon fields (Parisi-Sourlas mechanism) - to these particles. This was carried out in a series of papers \cite{Magpantay5}, \cite{Magpantay6}, \cite{Magpantay7}. In this work, I will restate the main points made in these papers. The main points made in the original papers are essentially correct except for a main correction. In Section VI, I will explicitly show gluon confinement and the mass gap in the effective action for gluons. This was missed in \cite{Magpantay5}. The correction to \cite{Magpantay5} essentially breaks translation symmetry, which at first glance seems wrong for it breaks a long believed symmetry of physical theories. But the breakdown of translation symmetry makes sense because of confinement, we really cannot translate the physics below the confinement regime where asymptotic freedom holds and above the confinement scale where the gluons and quarks do not exist but mesons and baryons of pion-nucleon physics and flavor dynamics. The correction will eventually lead to the proof of confinement of the gluons, the mass spectrum of the relevant vector degrees of freedom, i.e., the mass gap. 

When we add quarks in Section VII, we show the explicit confinement of dynamical quarks and a linear potential between these quarks by deriving a non-local four Fermi interaction with a linear potential. To arrive at the effective action for the degrees of freedom in the confining region, we make use of a known result in noise averaging. In summary, Section VII gives the effective action for the degrees of freedom in the confining regime - a set of Abelian vector fields that exhibit a mass gap and explicit confinement, fermions that are confined, interact locally with the Abelian vector fields and has a non-local, four-fermi interaction with linear potential. 

\section{\label{sec:level2}The Non-linear Gauge}
	I summarize the main points defined by equation (1). It contains (a) the Coulomb gauge $ \partial \cdot A = 0 $ and (b) the quadratic condition $ \partial \cdot A = f(x) \neq 0 $ with $ f(x) $ a zero mode of $ \partial \cdot D $. These two regimes do not mix in the sense that configurations that satisfy one cannot be gauge transformed to the other. 

Consider field configurations $ A(x) $ that satisfy the quadratic condition given by equation (1), with $ \partial \cdot A \neq 0 $. The operator $ \partial \cdot D $ is singular with  $ \partial \cdot A $ as the zero mode. Suppose we want to transform this field configuration to the Coulomb gauge, then we have to solve for a gauge parameter $ \Lambda(x) $ that satisfies
\begin{subequations}\label{2}
\begin{gather}
A^{'}(x)=A(x)+D(A)\Lambda(x),\label{first}\\
[\partial \cdot D(A)]\Lambda=-\partial \cdot A,
\end{gather}
\end{subequations}
where the second equation follows from the fact that we impose $ \partial \cdot A' =0 $. For $ \Lambda \neq 0 $ to exist, the zero mode of $ \partial \cdot D(A) $, which is $ \partial \cdot A $ must be orthogonal to the source in equation (2b). Obviously, this is not possible as the source is also the zero mode, thus we cannot gauge transform $ A $ to the Coulomb gauge. 

Another way to see this is by multiplying both sides of equation (2b) by $ \partial \cdot A $ and integrating both sides over $ R^{4} $. Integrating by parts at the left side, dropping the surface terms and using the fact that $ [\partial \cdot D(A)](\partial \cdot A) = [D(A)\cdot \partial](\partial \cdot A) $, we find the left hand side equals to zero while the right hand side, being an integral of a positive definite function is not zero. Thus, a gauge parameter $ \Lambda(x) $ does not exist to transform A to the Coulomb gauge.

The converse is also true. Suppose we start with a field configuration that satisfies the Coulomb gauge, i.e., $ \partial \cdot A = 0 $ and want to transform it to an A' that satisfies the quadratic gauge condition defined by (1) with $ \partial \cdot A' \neq 0 $. This means we have to solve for a gauge parameter $ \Lambda(x) $ such that
\begin{subequations}\label{3}
\begin{gather}
A' = A+D(A)\Lambda, \label{first}\\
\partial \cdot A' = [\partial \cdot D(A)]\Lambda,\label{second}\\
[\partial \cdot D(A')](\partial \cdot A') =0.
\end{gather}
\end{subequations}
Imposing the non-linear gauge fixing defined by equation (1) on A', we find $ \Lambda(x) $ must be solved from
\begin{equation}\label{4}
[\partial \cdot D(A)]^{2}\Lambda=0.
\end{equation}
We dropped the quadratic term in $\Lambda$ because transformations are infinitesimal. Let us take the case where $det(\partial \cdot D(A))\neq 0$, i.e., the operator is non-singular, thus it does not have a zero mode. But this contradicts equation (4), which says the operator $\partial \cdot D(A)$ has a zero mode which is precisely $\partial\ \cdot D(A)\Lambda$. Thus, we cannot gauge transform an A that satisfies the Coulomb gauge and has $det(\partial \cdot D(A)) \neq 0$, i.e., field configurations that are in the central Gribov region, to the non-linear gauge.

Suppose we have $det(\partial \cdot D(A))= 0$, the field A is on the Gribov horizon bounding the central Gribov region. Then equation (4) is naively consistent, i.e., $\partial \cdot D(A)$ having a zero mode, which is 
\begin{equation}\label{5}
\partial \cdot D(A)\Lambda=F(x). 
\end{equation}
But for $\Lambda $ to be solved from equation (5), the zero mode of $ \partial \cdot D(A) $, which is F(x), must be orthogonal to the source, which is also F(x). Thus, contradictory and therefore $ \Lambda $ does not exist and the field configurations on the Gribov horizon cannot be gauge transformed to the non-trivial part of the non-linear gauge fixing defined by equation (1). 

Another way to see this is by multiplying equation (5) by F(x) and integrating over $ R^{4} $. Integrating by parts at the LHS, we will get an integrand $ [(D(A) \cdot \partial)F(x)]\Lambda $. Since $ \partial \cdot A=0 $, the $ \partial \cdot D(A)=D(A) \cdot \partial $ and using equation (4), we find the LHS equals zero while the RHS, being a positive definite integral is non-zero. Thus, a contradiction and therefore $ \Lambda $ does not exist.

The preceding arguments show that the two regimes of the gauge condition (1), the linear, which is the Coulomb, and the quadratic part are distinct and cannot be gauge transformed to each other. Thus, there is a need to generalize the Coulomb gauge to the non-linear gauge defined by equation (1).

Before we explore the consequences of the non-linear gauge, we next show what was claimed in \cite{Magpantay1} that the non-linear gauge with $ \partial \cdot D(A) $ defining the non-linearity from the Coulomb gauge can no longer be extended. In other words, the gauge condition $ [(\partial \cdot D(A)]^{n}(\partial \cdot A))=0 $ terminates at $ n=1 $. To see this, consider 
\begin{equation}\label{6}
[\partial \cdot D(A)]^{2}(\partial \cdot A)=0.
\end{equation}
This condition has 3 regimes (a) $ \partial \cdot A = 0 $, (b) $ \partial \cdot A = f(x) \neq 0 $ and $ (\partial \cdot D(A))f(x) = 0 $, and (c) $ \partial \cdot A = f(x) \neq 0 $, $ (\partial \cdot D(A))f(x) = g(x) \neq 0 $ and $ (\partial \cdot D(A))g(x) = 0 $. The first is the Coulomb gauge, the second is the non-linear gauge defined by equation (1), while the third is the new regime. However, this regime is not consistent. The third regime says the operator $ \partial \cdot D(A) $ is singular with g(x) as its zero mode. But to solve for f(x) from $ (\partial \cdot D(A))f(x) = g(x) $, we must have the zero mode g(x) orthogonal to the source which is also g(x). Thus, not possible and the third regime is inconsistent. We can also arrive at this inconsistency by multiplying $ (\partial \cdot D(A)f = g(x) $ by g(x) and integrating in $ R^{4} $. Note that $ [\partial \cdot D(A)]f(x) = [D(A) \cdot \partial]f(x) $ since $ f(x) = \partial \cdot A $. Then integrating by parts, dropping of the surface term, we will get the LHS equals zero since $ (\partial \cdot D(A))g(x) = 0 $ while the RHS is not zero since it is an integral of a positive definite term, giving an inconsistency. 

What we have shown is that the non-linear generalization of the Coulomb gauge given by equation (1) can no longer be extended to higher order in A. A consistent non-linear generalization of the Coulomb gauge terminates at the quadratic level.

How does the Yang-Mills configuration space look like as defined by the gauge condition (1)? First, it contains the Coulomb gauge so we have the plane (since linear in A) defined by $ \partial \cdot A = 0 $. As Gribov \cite{Gribov} argued, this surface contains the central Gribov region labelled $ C_{0} $, where the operator $ \partial \cdot D(A) $ only has positive eigenvalues. This must be the perturbative regime. The boundary of the central Gribov region is $\textit{l}_{1} $, where the operator $ \partial \cdot D(A) $ has one zero eigenvalue, thus the existence of a copy. Non-perturbative physics supposedly follow from properly taking into account the copy. The next region is $ C_{1} $, where $ \partial \cdot D(A) $ has one negative eigenvalue and the boundary to that is $ \textit{l}_{2} $, which now has two zero eigenvalues. The process continues but we will only focus in regions $ C_{0} $ and $\textit{ l}_{1} $. 

Now consider the surface parallel to the Coulomb surface but infinitesimally displaced defined by $ \partial \cdot A = f(x) $. Equation (1) defines the first Gribov horizon of the surface $ \partial \cdot A = f(x) $, where the zero mode of $ \partial \cdot D(A) $ is f(x) itself. This horizon is labelled $ \textit{l}_{f1} $. This bounds the region $ C_{f0} $ and the points inside this region can be gauge transformed to points in $ C_{0} $. But as discussed above, the points in the horizon $ \textit{l}_{f1} $ is not gauge transformable to $ C_{0} $ and $ \textit{l}_{1} $. As we vary f(x), say to f'(x), which is infinitesimally different from f(x), we will be considering another parallel surface to the Coulomb surface and we will get another Gribov horizon $ \textit{l}_{f'1} $. How are the two horizons related. The answer is none because each horizon  $ \textit{l}_{f1} $ forms a gauge orbit as we will show below. 

We already argued above that fields in $ \textit{l}_{f1} $ cannot be gauge transformed to the Coulomb surface. These fields cannot also be gauge transformed to fields in $ C_{f0} $ because $ det(\partial \cdot D(A)) $ is invariant under infinitesimal transformation and this determinant is zero in  $ \textit{l}_{f1} $ and non-zero in $ C_{f0} $. Can we transform A in $ \textit{l}_{f1} $ to another A' that is also in $ \textit{l}_{f1} $. The answer is yes and the gauge parameter of the transformation is $ f(x) $ and no other because the parameter must satisfy $ [\partial \cdot D(A)]\Lambda(x) = 0 $ and in $ \textit{l}_{f1} $ there is only one zero mode and it is f(x). Thus all fields in $ \textit{l}_{f1} $ belong to one orbit of gauge transformations. 

How about gauge transforming a field A(x) in $ \textit{l}_{f1} $ to another surface A' defined by $ \partial \cdot A' = f'(x) $? Since $ det(\partial \cdot D(A')) $ in $ C_{f'0} $ is non-zero, there is no gauge transformation from A in $ \textit{l}_{f1} $ to A' in $ C_{f'0} $. How about to A' in  $ \textit{l}_{f'1} $? This is not obviously disallowed because both have zero determinants for their respective $ \partial \cdot D $. But there is no such transformation as we argue below.

The gauge parameter must be solved from
\begin{subequations}\label{7}
\begin{gather}
A' = A + D(A)\Lambda,\label{first}\\
[\partial \cdot D(A)](\partial \cdot A ) = 0,\label{2}\\
[\partial \cdot D(A')](\partial \cdot A') = 0.
\end{gather}
\end{subequations}
Using equation(7a) in (equation(7c) and making use of equation (7b), the gauge parameter $ \Lambda $ must be solved from
\begin{subequations}\label{8}
\begin{gather}
\Theta^{ab}\Lambda^{b} = 0,\label{first}\\
\Theta^{ab} = (D \cdot \partial)^{ac}(\partial \cdot D)^{cb} - g \varepsilon^{acd} \partial (\partial \cdot A^{c}) \cdot (D^{db}),
\end{gather}
\end{subequations}
the gauge covariant operator D in equation (8b) are all functions of A. As was shown in \cite{Magpantay1} and summarized here in Appendix A, the operator $ \Theta $ is non-singular even though $ \partial \cdot D(A) $ is singular. Thus equation (8b) only has the trivial solution $ \Lambda = 0 $ and there is no gauge transformation from A in $ \textit{l}_{f1} $ to A' in $ \textit{l}_{f'1} $. The field configurations in $ \textit{l}_{f1} $ is an orbit, i.e., all fields are gauge copies of each other.    

Taking all these into account, the Yang-Mills configuration space is represented by Figure 1. The meshed surface on $ \partial \cdot A = 0 $ represents the central Gribov region $ C_{0} $ and it is the perturbative regime. The boundary of $ C_{0} $ is the first Gribov horizon $ \textit{l}_{1} $. This is the linear part of the gauge condition defined by equation (1).  The quadratics part of equation (1) is represented by the paraboloid-ellipsoid. For a given $ \partial \cdot A = f(x) $, the boundary to $ C_{f0} $ given by $ \textit{l}_{f1} $ is one orbit. As we vary $ f $, we will generate the other boundaries $ \textit{l}_{f'1} $, which are separate orbits. Thus, the paraboloid-ellipsoid is a collection of orbits that are boundaries. If we integrate the contributions of all gauge potentials on this paraboloid-ellipsoid, there will be a massive over counting of gauge equivalent fields. Thus, we see that there is a need to look for a new way to express the potential to avoid this over counting. And one way is to simply integrate over the curve $ \Gamma $, which cuts through each orbit once. But then the count of degrees of freedom is not right, $ A_{\mu}^{a} $ has $ 3 x 4 $ degrees of freedom at each point while $ f^{a} $ only has 3 degrees of freedom. This will be settled in the next section.    
\begin{figure}[h]
\includegraphics[width=15 cm,height=12 cm]{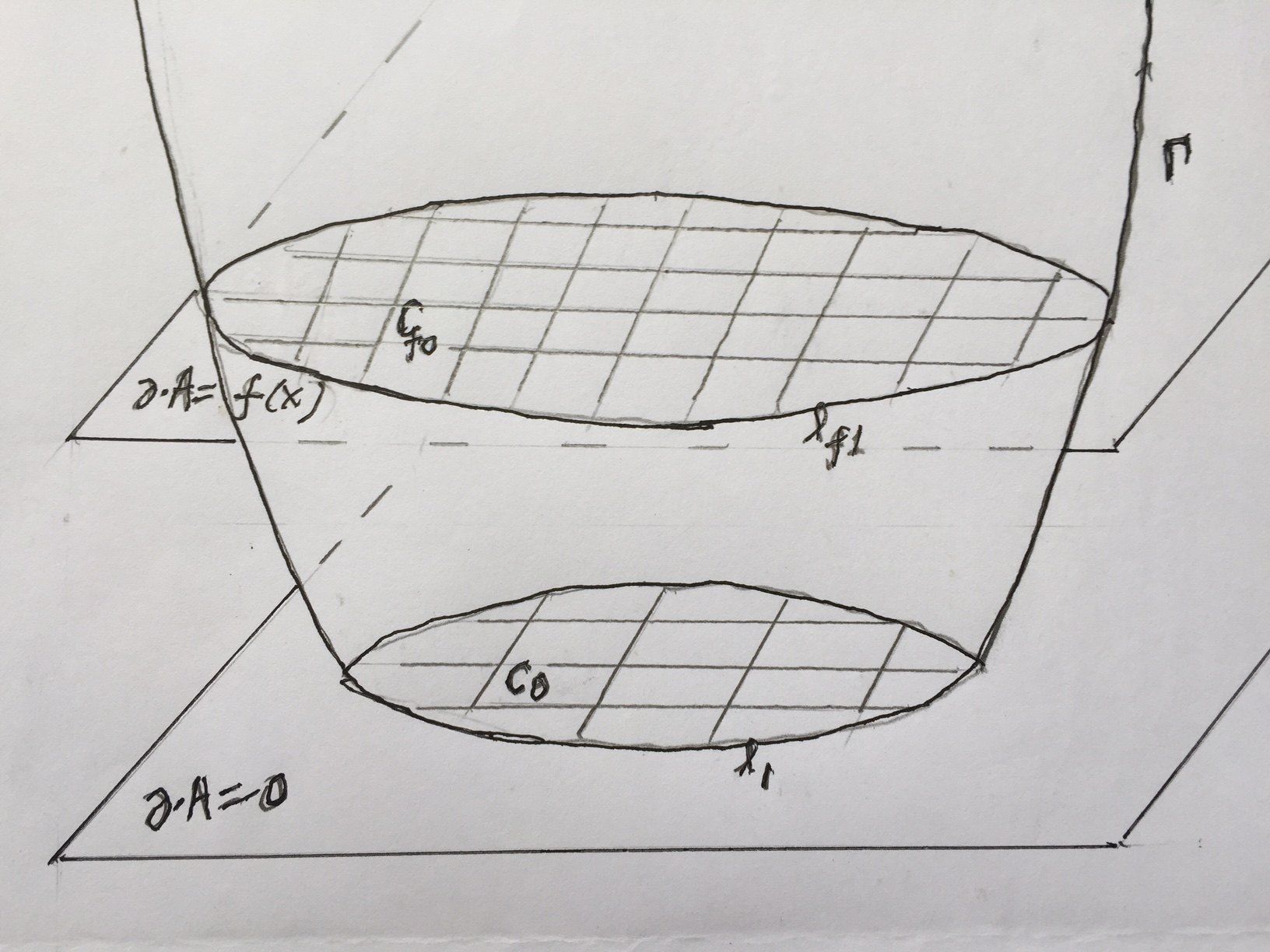}
\end{figure}

Figure 1. The Yang-Mills configuration space showing the linear and quadratic components of the gauge condition defined by equation 1.

\section{\label{sec:level3}The New Degrees of Freedom}
	We will essentially follow \cite{Magpantay3} and work in 4D Euclidean space $ R^{4} $ where the D'Alembertian operator is a 4D Laplacian, thus an elliptic operator. The gauge fixing condition I propose is written as 
\begin{subequations}\label{9}
\begin{gather}
\partial \cdot A^{a} = f^{a}, \label{first}\\
[\partial \cdot D^{ab}]f^{b} = 0,
\end{gather}
\end{subequations}
where we consider the gauge symmetry SU(2) and a, b runs from 1 to 3. These two equations are solved by 
\begin{equation}\label{10}
\textit{l}^{2}D_{\mu}^{ab}f^{b} = \left( A_{\mu}^{a} - \partial_{\mu} \dfrac{1}{\square}f^{a} \right) - t_{\mu}^{a},
\end{equation}
where a length scale $ \textit{l} $ is introduced for dimensional reasons ($ f^{a} $ has dimension of $ (mass)^{2} $) and $ t_{\mu}^{a} $ is transverse. Equation (10) is solved by 
\begin{equation}\label{11}
A_{\mu}^{a} = \dfrac{1}{(1+g^{2}\textit{l}^{4}\vec{f} \cdot \vec{f})}\left( \delta^{ab} + g\textit{l}^{2}\epsilon^{abc}f^{c} + g^{2}\textit{l}^{4}f^{a}f^{b} \right)\left( \textit{l}^{2}\partial_{\mu}f^{b} + \partial_{\mu}\frac{1}{\square}f^{b} + t_{\mu}^{\mu} \right).
\end{equation}
Rescaling the fields by $ g\textit{l}^{2}f^{a} \rightarrow f^{a} $ so that the new field $ f^{a} $ is dimensionless and defining a new vector field $ t_{\mu}^{a} + \dfrac{1}{g\textit{l}^{2}}\partial_{\mu}\frac{1}{\square}f^{a} \rightarrow t_{\mu}^{a} $, we find 
\begin{equation}\label{12}
A_{\mu}^{a} = \dfrac{1}{(1 + \vec{f}\cdot\vec{f})}\left( \delta^{ab} + \epsilon^{abc}f^{c} + f^{a}f^{b} \right) \left( \frac{1}{g}\partial_{\mu} f^{b} + t_{\mu}^{b} \right).
\end{equation}
Equation (12) says the 12 $ A_{\mu}^{a} $ is written down in terms of 3 $ f^{a} $ and 12 $ t_{\mu}^{a} $, a seeming over specification. But this is not so because the following constraints must hold 
\begin{subequations}\label{13}
\begin{gather}
\partial_{\mu} t_{\mu}^{a} = \frac{1}{g\textit{l}^{2}} f^{a},\label{first}\\
\rho ^{a} = \dfrac{1}{(1+\vec{f}\cdot\vec{f})^{2}}\left( -(\vec{f}\cdot\vec{f})\epsilon^{abc} + \epsilon^{abd}f^{d}f^{c} - \epsilon^{acd}f^{b}f^{d} + f^{a}\epsilon^{bcd}f^{d} \right)\partial_{\mu}f^{b}t_{\mu}^{c} = 0.
\end{gather}
\end{subequations}
With these constraints on $ f^{a} $ and $ t_{\mu}^{a} $, the number of degrees of freedom tallies with those of the original $ A_{\mu}^{a} $. The decomposition given by equation (12) shows the other degrees of freedom that go with each $ f^{a} $. 

	Following \cite{Magpantay4} the path integral of pure Yang-Mills in the non-linear gauge expressed in terms of the new degrees of freedom is 
\begin{equation}\label{14}
PI = \int (df^{a})(dt_{\mu}^{a}) \delta\left( \partial \cdot t^{a} - \frac{1}{g\textit{l}^{2}}f^{a}\right) \delta(\rho^{a}) [det(1+\vec{f}\cdot\vec{f})]^{-4} det(\Theta) \exp (-S_{YM}(f^{a},t_{\mu}^{a})),
\end{equation}
where $ det(\Theta) $ is the Fadeev-Popov determinant of the gauge condition defined by equation (1) and $ \Theta $ is given by equation (8b). Note that since $ \Theta $ is non-singular, $ det(\Theta) $ is never zero and we will not have a Gribov problem. To express the Yang-Mills action in terms of the new degrees of freedom, we express the field strength as \cite{Magpantay6}
\begin{equation}\label{15}
F_{\mu\nu}^{a} = \frac{1}{g}Z_{\mu\nu}^{a}(f) + L_{\mu\nu}^{a}(f,t) + gQ_{\mu\nu}^{a}(f,t),
\end{equation}
where Z does not depend on t, L is linear in t and Q is quadratic in t. The explicit expressions are 
\begin{subequations}\label{16}
\begin{gather}
Z_{\mu\nu}^{a}(f) = X^{abc}(f)\partial_{\mu}f^{b}\partial_{\nu}f^{c},\label{first}\\
L_{\mu\nu}^{a}(f,t) = R^{ab}(f)(\partial_{\mu}t_{\nu}^{b} - \partial_{\nu}t_{\mu}^{b}) + Y^{abc}(f)(\partial_{\mu}f^{b}t_{\nu}^{c} - \partial_{\nu}f^{b}t_{\mu}^{c}),\label{second}\\
Q_{\mu\nu}^{a}(f,t) = T^{abc}(f)t_{\mu}^{b}t_{\nu}^{c}.
\end{gather}
\end{subequations}
The f matrices are given by
\begin{subequations}\label{17}
\begin{gather}
R^{ab}(f) = \dfrac{1}{(1+\vec{f}\cdot\vec{f})}(\delta^{ab} + \epsilon^{abc}f^{c} + f^{a}f^{b}),\label{first}\\
\begin{split}
X^{abc}(f)& = \dfrac{1}{(1+\vec{f}\cdot\vec{f})^{2}}[-(1+2\vec{f}\cdot\vec{f})\epsilon^{abc} + 2\delta^{ab}f^{c} - 2\delta^{ac}f^{b} \\ 
&\quad + 3\epsilon^{abd}f^{d}f^{c} - 3\epsilon^{acd}f^{d}f^{b} + \epsilon^{bcd}f^{a}f^{d}],
\end{split}\label{second}\\
\begin{split}
Y^{abc}(f)& = \dfrac{1}{(1+\vec{f}\cdot\vec{f})^{2}}[-(\vec{f}\cdot\vec{f})\epsilon^{abc} + (1+\vec{f}\cdot\vec{f})f^{a}\delta^{bc} - (1-\vec{f}\cdot\vec{f})\delta^{ac}f^{b} \\
&\quad + 3\epsilon^{cad}f^{d}f^{b} - 2f^{a}f^{b}f^{c} + \epsilon^{abd}f^{d}f^{c} + f^{a}\epsilon^{bcd}f^{d}],\end{split} \label{third}\\
\begin{split}
T^{abc}(f)& = \dfrac{1}{(1+\vec{f}\cdot\vec{f})^{2}}[\epsilon^{abc} + (1+\vec{f}\cdot\vec{f})f^{b}\delta^{ac} - (1+\vec{f}\cdot\vec{f})f^{c}\delta^{ab} \\
&\quad + \epsilon^{abd}f^{d}f^{c} + f^{a}\epsilon^{bcd}f^{d} + \epsilon^{acd}f^{d}f^{b}].
\end{split}
\end{gather}
\end{subequations}
The action is still quartic in $ t_{\mu}^{a} $ but infinitely non-linear in $ f^{a} $. Looking for classical solutions that may have physics consequences is now rather difficult. The way we will explore this is by first looking at the most non-linear dynamics, which is that of the $ f^{a} $ and then look at how its classical solution, which we will call $ \tilde{f}^{a} $, influence the $ t_{\mu}^{a} $ dynamics and later also the quark dynamics.

\section{\label{sec:level4}The Significance of the Isoscalar field}
	To motivate the significance of the isoscalar field $ f^{a}(x) $, recall what we discussed already. In electrodynamics, the transverse gauge condition (Minkowski, 3+1 space-time) $ \triangledown \cdot \vec{A} = 0 $ exposes the physical degrees of freedom, the transverse photons. In the short distance regime of Yang-Mills, the running coupling goes to zero, and the (massless) transverse gluons are physical degrees of freedom. But as we increase the distance scale, the coupling becomes large and the massless gluons are replaced by new degrees of freedom that exhibit a mass gap. The new massive degrees of freedom, which only exists well inside the hadrons, thus confined, cannot be transverse. This is what we will show in Section VI, which is based on the discussion in this section.
	
	In the configuration space of the gauge potentials, these fields are accounted for by the surfaces parallel to the  $ \partial \cdot A^{a} = 0 $ surface. And fixing the gauge using equation (1) is tantamount to choosing the new vector fields $ t_{\mu}^{a} $ for every $ f^{a} $ chosen. The new fields to represent the gauge fields however must satisfy the constraints  
defined by equations (13a) and (13b). 
	
	Consider the most non-linear part of the action, the pure $ f^{a}(x) $, term of the Yang-Mills field strength. The field equations are too non-linear to be solved. But there is one obvious solution and that is
\begin{equation}\label{18}
Z_{\mu\nu}^{a} = X^{abc}\partial_{\mu}f^{b}\partial_{\nu}f^{c} = 0.
\end{equation}
From equation (17b), we notice that $ X^{abc} $ is anti-symmetric with respect to the last two indices. This suggests that if $ f^{a}(x) = f^{a}(\vert x \vert) $ with $ \vert x \vert = (x_{\mu}x_{\mu})^{\frac{1}{2}} $, i.e., purely a function of the radial coordinate in $ R^{4} $, then it is a classical solution. This means all $ \tilde{f}^{a}(\vert x \vert) $ are classical solutions. This must be a dense set of classical solutions and we need a way of handling all these solutions.

	In \cite{Magpantay3}, I proposed that  $ \tilde{f}^{a}(\vert x \vert) $ be treated as a white noise. This means the interaction with a specific white noise $ \tilde{f}^{a}(\vert x \vert) $ is not important. What is important is the contribution of all the white noises to quarks and gluons and this necessitates an averaging procedure. The white noise distribution is given by
\begin{equation}\label{19}
P[\tilde{f}] = \textit{N}\exp {(-\frac{1}{\textit{l}}\int_{0}^{\infty} ds \tilde{f}^{a}(s)\tilde{f}^{a}(s))},
\end{equation}
where $ \textit{N} $ is the normalization whose explicit form will be given when we discretize the integral. Note since $ \tilde{f}^{a} $ is dimensionless and s has dimension of length, we introduced  the length scale $ \textit{l} $.  
 
	The explicit averaging over the gaussian white noise $ \tilde{f}^{a}(x) $ to get the effective dynamics of quarks and 'gluons' will be done in Sections V, VI and VII. For the moment, we take note of a seeming problem if we identify the $ \tilde{f}^{a}(\vert x \vert) $ as the ground state. Since this classical configuration has $ Z_{\mu\nu}^{a} = 0 $, thus giving $ (Z_{\mu\nu}^{a}Z_{\mu\nu}^{a})_{vac} = 0 $, which is in apparent contradiction to the fact that the QCD vacuum is characterized by non-zero condendates, in particular, gluon condensate \cite{Shifman}. But there is no contradiction as can be seen from equation (15) because the ground state will not just be determined by the $ f^{a}(x) $ dynamics but also by the 'gluons' defined by $ t_{\mu}^{a}(x) $. The resolution will be done in Section VI.
	
	The discussions in the next sections will show that the mechanism for non-perturbative physics of QCD - mass gap and quark and 'gluon' confinement - is the stochastic treatment of the isoscalar field $ f^{a}(x) $. 	 
	
\section{\label{sec:level5}Hints of Confinement}   	     
	Confinement must be shown for dynamical quarks and gluons. But this is a more formidable problem so we will begin with hints of confinement that follow from the non-linear gauge defined by equation (1). In this section, we will just review the results of \cite{Magpantay3} and \cite{Magpantay4} as we find that the results are essentially correct. 
	
	We begin by calculating the gluon propagator $ \left\langle 0 \vert A_{\mu}^{a}(x)A_{\nu}^{b}(y) \vert 0 \right\rangle $. Starting from $ \partial \cdot A = \dfrac{1}{g\textit{l}^{2}} f $ and taking f to be the random classical solution $ \tilde{f} $, we have
\begin{equation}\label{20}
\partial_{\mu}^{x}\partial_{\nu}^{y} \left\langle 0 \vert A_{\mu}^{a}(x)A_{\nu}^{b}(y) \vert 0 \right\rangle = \dfrac{1}{g^{2}\textit{l}^{4}} \tilde{f}^{a}(\vert x \vert)\tilde{f}^{b}(\vert y \vert).
\end{equation}  		
Taking the average of this equation using the distribution given by equation (19), we find the right hand side is equal to $ \textit{l} \delta^{ab} \delta(\vert \vert x \vert - \vert y \vert \vert) $. Since the RHS is purely a function of the radial coordinates, the gauge potentials must also have the same dependence. If this is the case, then $ \partial_{\mu} = \frac{x_{\mu}}{x}\frac{d}{dx} $. It then follows that
\begin{equation}\label{21}
\left\langle 0 \vert A_{\mu}^{a}(\vert x \vert)A_{\nu}^{b}(\vert y \vert) \vert 0 \right\rangle = \dfrac{1}{g^{2}\textit{l}^{3}} \delta_{\mu\nu}\delta^{ab} \vert \vert x \vert - \vert y \vert \vert.
\end{equation}
Note that the angle that goes with $ x \cdot y $ is not important as we can always choose the origin so that x and y are collinear as we deal with two point function above. Equation (21) gives the linear potential in 3D when we take $ x_{0} $ and $ y_{0} $ both go to zero in the instantaneous limit.

	The next hint of confinement we will cite is the area law behaviour of the Wilson loop, which implies a linear potential in 3D for static quarks. The Wilson loop is given by
\begin{equation}\label{22}
W = Tr P \exp {(i \oint_{\Gamma} A \cdot dx)},
\end{equation}  	
where the path of integration defined by $ \Gamma $ is a rectangular path in a 1-space (x axis) vs time (y axis) Cartesian plot.
	
The vector field $ A_{\mu}^{a}(x) $ is given by equation (12) with $ t_{\mu}^{a} = 0 $. In taking the stochastic average of the Wilson loop, we have to carefully consider the points in figure 1 that have the same Euclidean length because the distribution given by equation (19) is a function only of the invariant Euclidean length. Furthermore, we have to discretize the distribution as 
\begin{equation}\label{23}
P[\tilde{f}] = \textit{N}\exp {(-\sigma \sum_{s,a} \tilde{f}^{a}(s)\tilde{f}^{a}(s))},
\end{equation}
where $ \sigma = \dfrac{\vartriangle s}{l_{0}} $ and $ \textit{N} = \displaystyle\prod_{s} (2\pi^{3})^{\frac{-1}{2}}\sigma^{\frac{3}{2}} $. Note that as we let $ \vartriangle s \rightarrow 0 $, $ \sigma \rightarrow 0 $. When we do the averaging at a particular point s, we take note that the 'volume' $ \displaystyle\prod_{a}d\tilde{f}^{a}(s) = r^{2}dr d\Omega $, where $ r^{2} = \tilde{f}^{a}(s)\tilde{f}^{a}(s) $ and $ d\Omega $ is the solid angle in SU(2) space with total solid angle of $ 8\pi $.    
We use the following integral from Gradshteyn and Ryzhik page 338 \cite{Gradshteyn}
\begin{subequations}\label{24}
\begin{gather}
I(\sigma,\beta^{2}) = \int_{0}^{\infty} \dfrac{\exp {-\sigma r^{2}}}{r^{2}+\beta^{2}} dr,\label{first}\\
I(\sigma,\beta^{2}) = [1 - \Phi(\sigma^{\frac{1}{2}}\beta)]\frac{\pi}{2\beta}\exp {\beta^{2}\sigma},
\end{gather}
\end{subequations}
valid for $ Re \beta > 0 $ and $ arg \sigma < \frac{\pi}{4} $. The function $ \Phi $ is the error function given by
\begin{equation}\label{25}
\Phi(\beta \sigma^{\frac{1}{2}}) = \frac{2}{\sqrt{\pi}} \int_{0}^{\beta \sigma^{\frac{1}{2}}} \exp {(-t^{2})} dt.
\end{equation}
For our purposes, what is useful is the series expansion of the error function and it is given by
\begin{equation}\label{26}
\Phi(\sigma \beta^{2}) = \frac{2}{\sqrt{\pi}} \sum_{k=1}^{\infty} \dfrac{(-1)^{k+1}}{(2k-1)(k-1)!} (\sigma \beta^{2})^{\dfrac{2k-1}{2}},
\end{equation}
where we considered $ \Phi $ to be a function of $ \sigma \beta^{2} $ instead of $ \beta \sigma^{\frac{1}{2}} $ because the integrals we will meet follow from differentiating I of equation (24) with $ \beta^{2} $ and $ \sigma $. To calculate the stochastic average of the Wilson loop, eventually we have to put $ \beta^{2} \rightarrow 1 $ and $ \sigma \rightarrow 0 $. Following the discussions in \cite{Magpantay3}, we find that the stochastic average of the Wilson loop goes like 
\begin{equation}\label{27}
\left\langle W(\tilde{f}) \right\rangle = \exp {(-\dfrac{1}{g^{2}\triangle x \triangle t}LT)}.
\end{equation}
Note the area law behaviour. In the above equation, $ \triangle x $ and $ \triangle t $ are the lattice spacings.

	The final hint of confinement we discuss is showing that a Parisi-Sourlas mechanism \cite{Parisi} can be had also for Yang-Mills theory. As a reminder, Parisi and Sourlas showed dimensional reduction, from 4 to 2 dimensions, for scalar fields coupled to random fields. The dimensional reduction follows from the fact that a stochastic supersymmetry follows from such a system and since fermionic coordinates have negative dimension, the theory was shown to be equivalent to a similar scalar field in 2 dimensions less. This looks like a compelling idea in showing confinement except that it has two drawbacks. The first is that Yang-Mills involves vector fields so what happens to the other components of the vector fields when we lose 2 dimensions. Second, there are no fundamental scalar fields in QCD so how do we show dimensional reduction for scalars in QCD when there is none. 
	
	The non-linear regime of the gauge condition (1) led to a decomposition of the gauge potential given by equation (12) and the field strength given by equations (15) and (16). Right away we see the existence of a scalar field and if we put $ t_{\mu}^{a} = 0 $, the action of pure $ f^{a} $ is quartic in derivatives and infinitely non-linear as equation(16a) shows. From the path-integral given by equation (14) (with $ t_{\mu}^{a} = 0 $ it is not at all obvious that the theory is equivalent to a non-linear sigma model. But it is as the rather involved derivation in \cite{Magpantay5} showed. In the following, we will just reiterate the main points followed in this paper.

	My starting point in \cite{Magpantay4} for the pure $ f^{a} $ path-integral coming from equation (14) to show the Parisi-Sourlas mechanism is
\begin{equation}\label{28}
PI(f) = \int (df^{a}) [det(1+\vec{f}\cdot\vec{f})]^{-1} det(\Theta) \exp {(-\frac{1}{g^{2}}\int \frac{1}{4}Z^{2})},
\end{equation}
where $ \Theta $ is the operator given by equation (8) with $ t_{\mu}^{a} = 0 $, and we changed the power of $ det(1+\vec{f}\cdot\vec{f}) $ from -4 to -1 by changing the power of $ (1+\vec{f}\cdot\vec{f})^{2}) $ in $ \rho ^{a} $ in equation (13b) from -2 to +1 in equation (14) before putting $ t_{\mu}^{a} = 0 $. This is wrong as I will show below and in the process derive the changes in my proof of the equivalence to a Parisi-Sourlas mechanism. 

	The error in simply putting $ t_{\mu}^{a} = 0 $ in equation (14) will result in $ \delta(f^{a}) $, thus everything is trivial, there is no Parisi-Sourlas mechanism to show. Go back to equation (14), include the $ det^{-3}(1+\vec{f}\cdot\vec{f}) $ in the $ \delta(\rho^{a}) $ resulting in a constraint equation of the form
\begin{equation}\label{29}
(1+\vec{f}\cdot\vec{f})^{3}\rho^{a} = M^{ab}_{\mu}(f)t_{\mu}^{b},
\end{equation}
where $ M^{ab}_{\mu}(f) $ can be read off equation (13b) and it has a factor $ (1+\vec{f}\cdot\vec{f}) $ factor instead of $ (1+\vec{f}\cdot\vec{f})^{-2} $. The constraint defined by equation (13b) involves 3 equations $ (a = 1, 2, 3) $ with 12 vector fields $ t_{\mu}^{a} $ and the matrix  $ M^{ab}_{\mu}(f) $ is a $ 3 X 12 $ matrix. Since
\begin{equation}\label{30}
\delta(M^{ab}_{\mu}t_{\mu}^{b}) = \dfrac{\delta(t_{\mu}^{a})}{det M^{ab}_{\mu}},
\end{equation}
where now we see how we can put $ t_{\mu}^{a} = 0 $. But the problem is M is not a square matrix, it is a $ 3 X 12 $ matrix. Fortunately, mathematicians showed how to compute such a determinant. Radic defined such a determinant \cite{Radic}, \cite{Amiri} by first defining 12 $ (3 X 1) $ column vectors from the $ 3 X 12 $ matrix M labelled by $ {M_{1},...., M_{12}} $. The determinant, following the notation in \cite{Makarewicz}, is then given by
\begin{equation}\label{31}
det(M) = \sum_{1\leq j_{1}<j_{2}<j_{3}\leq12} (-1)^{r+j_{1}+j_{2}+j_{3}}det(M_{j_{1}}M_{j_{2}}M_{j_{3}}).
\end{equation}
Note, the matrix $ (M_{j_{1}}M_{j_{2}}M_{j_{3}}) $ is a square $ (3 X 3) $ matrix and there are 55 such matrices whose determinants will be added or subtracted to determine the determinant of the $ 3 X 12 $ $ M^{ab}_{\mu} $ matrix. Since the elements of each of the 12 column vectors $ M_{i} $ are polynomial functions of $ f^{a} $ and $ \partial_{\mu}f^{a} $ up to triplic order (aside from the common $ (1+\vec{f}\cdot\vec{f}) $ factor, then each of the of the 55 determinant terms will be at most up order 9 in these polynomials. Thus, we can safely say that we can find a $ 3 X 3 $ matrix N that is also a polynomial function of $ f^{a} $ and $ \partial_{\mu}f^{a} $ such that $ det(M) = det(N) $.  
	
	Substituting this now in equation (30), which in turn is substituted in equation (14), we find the resulting path-integral after integrating out $ t_{\mu}^{a} $
\begin{equation}\label{32}
PI(f) = \int (df^{a}) [det(1+\vec{f}\cdot\vec{f})]^{-1} det(\Theta) \delta(N^{ab}f^{b}) \exp {(-\frac{1}{g^{2}}\int \frac{1}{4}Z^{2})},
\end{equation}
where we included $ det^{-1}(N) $ with the $ \delta(f^{a}) $ and just ignored the constant $ \frac{1}{g\textit{l}^{2}} $. Now we will exponentiate the delta function in terms of a gaussian, i.e., 
\begin{equation}\label{33}
\delta(N^{ab}f^{b}) = \exp {\left( -\frac{1}{\xi} \int d^{4}x [N^{ab}f^{b}]^{2} \right)},
\end{equation}
where the limit $ \xi \rightarrow 0 $ should be taken at the end. The claim is that this is equivalent to a non-linear sigma model with path-integral given by
\begin{equation}\label{34}
PI(f) = \int (df^{a}) det\left( \dfrac{\delta ^{2}S}{\delta f^{a}\delta f^{b}} \right) \exp {\left[ -\frac{1}{2}\int d^{4}x \left( \dfrac{\delta S}{\delta f^{a}} \right)^{2} \right]}, 
\end{equation}
where 
\begin{subequations}\label{35}
\begin{gather}
S = \frac{1}{2g}\int d^{4}x  \eta^{ab}  \partial_{\mu}f^{a}\partial_{\mu}f^{b},\label{first}\\
\eta^{ab} = -\delta^{ab} + \dfrac{f^{a}f^{b}}{(1+\vec{f}\cdot\vec{f})}.
\end{gather}
\end{subequations}
Equations (35a) and (35b) describe the O(1,3) sigma model in the non-linear form. If path-integral given by equation (34) with the action given by equation (35) is equivalent to the path-integral given by equation (32), then we show the equivalence of pure Yang-Mills in terms of pure $ f^{a} $ dynamics being equivalent to an O(1,3) sigma model in a random field.
	
	The proof still follows \cite{Magpantay4} starting from equation (32) of that paper, which should now read
\begin{equation}\label{36}
\frac{1}{2}\int d^{4}x \left( \dfrac{\delta S}{\delta f^{a}} \right)^{2} = \frac{1}{4}\int d^{4}x Z^{2} - \frac{1}{\xi}\int d^{4}x [N^{ab}f^{b}]^{2} + \int d^{4}x \partial_{\mu}H_{\mu},
\end{equation}
The total divergence in that paper, which is given by equation (33) of that paper will now be supplemented by 
\begin{equation}\label{37}
\partial_{\mu}\textbf{H}_{\mu} = +\frac{1}{\xi}[N^{ab}f^{b}]^{2},
\end{equation}
thus cancelling the additional term in (36).
This total divergence, just like the original divergence given by equation (33) in \cite{Magpantay4} also has vanishing surface term. The reason is simple, $ [Nf]^{2} $ is an eight order polynomial in f with some terms having derivatives in f also. Since $ f \sim \frac{1}{x^{1+\varepsilon}} $ and $ \varepsilon $ infinitesimal, as $ x \rightarrow \infty $ for square-integrability of $ A_{\mu}^{a} $,   equation (35) of \cite{Magpantay4} clearly shows the additional surface term is also zero. To complete the proof of equivalence of equation (32) to equation (34), we just need to show 
\begin{equation}\label{38}
det^{-1}(1+\vec{f}\cdot\vec{f})det(\Theta) \approx det(\dfrac{\delta^{2}S}{\delta f^{a}\delta f^{b}}).
\end{equation} 
The proof given in \cite{Magpantay4} is quite involved but I think is still valid.

Thus, we have shown that the pure $ f^{a} $ action is expressible as an O(1,3) sigma model in a non-linear form.

	The path-integral given by equations (34) and (35) can be written as
\begin{subequations}\label{39}
\begin{gather}
PI(f^{a};w_{a};\bar{\psi^{a}},\psi^{a}) = \int (df^{a})(dw_{a})(d\psi^{a})(d\bar{\psi}^{a}) \exp {(-A)},\label{first}\\
A = \int d^{4}x \left\lbrace -\frac{1}{2}w_{a}^{2} + w_{a}\dfrac{\delta S}{\delta f^{a}} + \bar{\psi}^{a} \dfrac{\delta^{2} S}{\delta f^{a}\delta f^{b}} \psi^{b} \right\rbrace. 
\end{gather}
\end{subequations}
Unfortunately, because of the metric $ \eta^{ab} $, the action A cannot be written in terms of the supersymmetrized version of S (see equation (35), i.e.,
\begin{equation}\label{40}
A \neq \int d^{4}x d\bar{\theta} d\theta \frac{1}{2} \eta^{ab}(\Phi)\left( \partial_{\mu}\Phi^{a}\partial_{\mu}\Phi^{b} + \partial_{\theta}\Phi^{a}\partial_{\bar{\theta}}\Phi^{b} \right),
\end{equation}
where the superfield is defined by
\begin{equation}\label{41}
\Phi^{a} = f^{a} + \bar{\theta}\psi^{a} + \bar{\psi}^{a}\theta + \bar{\theta}\theta w^{a}.
\end{equation}
Also because of the metric $ \eta^{ab}(\Phi) $, the supersymmetrized $ S(\Phi) $ is not invariant under the supersymmetry transformations
\begin{subequations}\label{42}
\begin{gather}
x_{\mu} \rightarrow x'_{\mu} = x_{\mu} + \varepsilon_{\mu}\bar{\rho} \theta + \varepsilon_{\mu} \bar{\theta} \rho,\label{first}\\
\theta \rightarrow \theta' = \theta -2 \rho \varepsilon \cdot x,\label{second}\\
\bar{\theta} \rightarrow \bar{\theta}' = \bar{\theta} - 2\bar{\rho} \varepsilon \cdot x,
\end{gather}
\end{subequations}
which leave $ x^2 + \bar{\theta}\theta $ invariant, an important component in proving dimensional reduction. This shows dimensional reduction cannot be proven using the non-linear action of O(1,3) sigma model.

	The paper of Parisi and Sourlas hints that we should use instead the linear version of the O(1,3) sigma model given by
\begin{equation}\label{43}
S_{\sigma} = \int d^{4}x \left\lbrace \frac{1}{2}(\partial_{\mu}\sigma)^{2} -\frac{1}{2}(\partial_{\mu}f^{a})^{2} + \lambda (\sigma^{2} - \vec{f}\cdot\vec{f} -1) \right\rbrace .
\end{equation}
After checking again the calculations in \cite{Magpantay4}, the path-integral given by equation (34) with equations (35a) and (35b) is indeed equal to 
\begin{subequations}\label{44}
\begin{gather}
PI = \int (d\phi_{i})(dw_{i})(d\bar{\psi}_{i})(d\psi_{i}) \exp {(-A_{ss})},\label{first}\\\begin{split}
A_{ss}& = \int d^{4}x \bigg\lbrace -\frac{1}{2}w_{a}^{2} -\frac{1}{2}w_{\sigma}^{2} + w_{a}\dfrac{\delta S_{\sigma}}{\delta f_{a}} \\ 
&\quad + w_{\sigma} \dfrac{\delta S_{\sigma}}{\delta \sigma} + w_{\lambda} \dfrac{\delta S_{\sigma}}{\delta \lambda} + \bar{\psi}_{i}\dfrac{\delta^{2}S_{\sigma}}{\delta \psi_{i}\delta \bar{\psi}_{j}} \psi_{j} \bigg\rbrace,
\end{split} 
\end{gather}
\end{subequations}
where the fields $ \phi_{i} = (f^{a}, \sigma, \lambda) $, $ w_{i} = (w^{a}, w_{\sigma}, w_{\lambda}) $ and $ \Psi_{i} = (\Psi^{a}, \Psi_{\sigma}, \Psi_{\lambda}) $. The point is that the supersymmetric action $ A_{ss} $ given by equation (44b) can be written down in terms of the superfield action given by
\begin{equation}\label{45}
\begin{split}
A_{ss}& = S_{\sigma}(\Phi^{a}, \Phi_{\sigma} , \Phi_{\lambda}) = \int d^{4}x d\bar{\theta}d\theta \bigg\lbrace \frac{1}{2}\partial_{\mu}\Phi_{\sigma}\partial_{\mu}\Phi_{\sigma} + \frac{1}{2}\partial_{\theta} \Phi_{\sigma} \partial_{\bar{\theta}} \Phi_{\sigma} \\
&\quad + \frac{1}{2}\partial_{\mu}\Phi^{a}\partial_{\mu}\Phi^{a} + \frac{1}{2}\partial_{\theta}\Phi^{a}\partial_{\bar{\theta}}\Phi^{a} + \Phi_{\lambda}\left[ \Phi_{\sigma}^{2} - \Phi^{a}\Phi^{a} - 1\right] \bigg\rbrace   ,
\end{split}
\end{equation}
where $ \Phi^{a} $ is given by equation (41) and 
\begin{subequations}\label{46}
\begin{gather}
\Phi_{\sigma} = \sigma + \bar{\theta}\psi_{\sigma} + \bar{\psi}_{\sigma}\theta + \bar{\theta}\theta w_{\sigma},\label{first}\\
\phi_{\lambda} = \lambda + \bar{\theta}\psi_{\lambda} + \bar{\psi}_{\lambda}\theta + \bar{\theta}\theta w_{\lambda}.
\end{gather}
\end{subequations}
In the form of equation (46), the non-linear O(4) sigma model in a stochastic background is explicitly invariant under the susy transformations given by equation (43) and dimensional reduction follows, i.e., we get the O(4) sigma model in 2D. Then we rotate back $ f^{a} \rightarrow if^{a} $ to get again the O(1,3) sigma model in 2D. Thus, we have shown that the pure $ f^{a} $ dynamics of Yang-Mills theory exhibits the Parisi-Sourlas mechanism.

\section{\label{sec:level6}Gluon Confinement and the Mass Gap} 
	This section corrects a significant error in \cite{Magpantay5}. The correction is significant - it leads to an effective action for the $ t_{\mu}^{a} $, which we identify as 'gluons', that breaks translation symmetry. Furthermore, we explicitly show 'gluon' confinement and show that the mass spectrum of the 'gluons' follow from satisfying boundary conditions at the confinement length scale and at the asymptotic freedom scale. 

	The hints of confinement we discussed above will confine static quarks (linear potential from propagator in white noise and Wilson loop) and dimensional reduction for the scalar field (the divergence of the gluon field) action. But as we emphasized already, the confined particles are the gluons and dynamical quarks. In this section, we show the confinement of the gluons.

	We begin with the pure Yang-Mills action and make use of equation (12), which results in the field strength given by equations (15), (16) and (17). Since all spherically symmetric functions $ \tilde{f}^{a}(\vert x \vert) $, are classical solutions of the pure $ f^{a} $ action, we introduce the distribution given by equation (19) to be able to average the contributions of each classical configuration $ \tilde{f}^{a}(\vert x \vert) $. In essence, we are treating each classical configuration as a white noise. We next propose the background decomposition
\begin{equation}\label{47}
f^{a}(x) = \tilde{f}^{a}(\vert x \vert) + \phi(x),
\end{equation}
Substituting equation (47) in equations (15) to (17), the action is infinitely non-linear in $ \phi^{a} $ but still quartic in $ t_{\mu}^{a} $ with coefficients that are functions of $ \tilde{f}^{a}(\vert x \vert) $. Each particular classical configuration $ \tilde{f}^{a}(\vert x \vert) $ will give an effective dynamics for $ \phi^{a} $ and $ t_{\mu}^{a} $. A meaningful way to account for all the classical configurations $ \tilde{f} $ is to average their contributions and I proposed before to make use of the distribution given by equation (19). In this case we want to evaluate $ \left\langle S(\tilde{f}^{a},\phi^{a},t_{\mu}^{a})\right\rangle _{\tilde{f}} $ given by
\begin{equation}\label{48}
\left\langle S \right\rangle _{\tilde{f}} = \frac{1}{4}\int d^{4}x \left\langle \left\lbrace \frac{1}{g^{2}}Z^{2}+\frac{2}{g}ZL+(2ZQ+LL)+2gLQ+g^{2}QQ \right\rbrace \right\rangle _{\tilde{f}}.
\end{equation}
Each of the terms in equation (48) is an infinite series in $ \phi^{a}(x) $ with coefficients that are functions of $ \tilde{f}^{a}(\vert x \vert) $. It is these functions that we average using the distribution given by equation (19).

	But before we can start evaluating the stochastic averages, we need to define the derivative of a white noise, which is an ill-defined quantity. A physicists definition is to smooth out the derivative by defining
\begin{subequations}\label{49}
\begin{gather}
\partial_{\mu}\tilde{f}^{a}(x) = \dfrac{x_{\mu}}{x}\dfrac{d\tilde{f}^{a}(x}{dx}\label{first}\\
\dfrac{d\tilde{f}^{a}(x)}{dx} = \dfrac{\tilde{f}^{a}(x+\frac{\textit{l}}{\alpha})-\tilde{f}^{a}(x)}{\frac{\textit{l}}{\alpha}},
\end{gather}
\end{subequations}
where we just used x in the denominators to represent the Euclidean length and $ \alpha $ is a number bigger than 1. This number will turn out to be related to the mass gap as we will show later in this section. This is a main correction to \cite{Magpantay5}, where we just put $ \alpha = n $, an integer without any justification except that the smoothing of the white noise derivative must be within the confinement length $ \textit{l} $.  All the integrals that we meet in the averaging are of the form
\begin{equation}\label{50}
I_{n}^{m}(\sigma,\beta^{2}) = \int_{0}^{\infty} dr \dfrac{r^{2m}}{(\beta^{2}+r^{2})^{n}} \exp {(-\sigma r^{2})}.
\end{equation}
Using equation (24a), we find
\begin{equation}\label{51}
I_{n}^{m}(\sigma,\beta^{2}) = \dfrac{1}{(n-1)!(-1)^{m+n-1}}\dfrac{\partial^{n-1+m}I(\sigma,\beta^{2})}{(\partial\beta^{2})^{n-1}(\partial\sigma)^{m}}.
\end{equation}
Next, we make use of the series expansion of the error function as given in equation (26) and taking note that we will eventually take $ \beta^{2} = 1 $ and $ \sigma \rightarrow 0 $, we find      		  \begin{equation}\label{52}
2^{\frac{-1}{2}}\pi^{\frac{-3}{2}}\sigma^{\frac{3}{2}}I_{n}^{m} = \begin{cases}
0, & \text{if $ m \leq n $ }, \\
finite, & \text {if $ m = n+1 $},\\
diverges, & \text {if $ m \geq n+2 $ }.
\end{cases}
\end{equation}
Equations (50) to (52) are needed in evaluating averages involving even numbers of $ \tilde{f} $. The averages over odd numbers of $ \tilde{f} $ are all zeroes.  
	We are now in a position to evaluate each term in equation (48). 
\begin{equation}\label{53}
\frac{1}{g^{2}}\left\langle Z_{\mu\nu}^{a}(\tilde{f},\phi)Z_{\mu\nu}^{a}(\tilde{f},\phi) \right\rangle_{\tilde{f}} = 0
\end{equation}
All the terms, when expanding Z in terms of $ \phi $, satisfy the first of equation (46). The next is
\begin{equation}\label{54}
\frac{2}{g}\left\langle Z_{\mu\nu}^{a}(\tilde{f},\phi)L_{\mu\nu}^{a}(\tilde{f},\phi,t_{\mu}^{a}) \right\rangle _{\tilde{f}} = 0.
\end{equation}
Again, all the averages are of the form in the first of equation (52). There is no linear term in $ t_{\mu}^{a} $. The term triplic in $ t_{\mu}^{a} $ is also zero, i.e.,
\begin{equation}\label{55}
2g\left\langle L_{\mu\nu}^{a}(\tilde{f},\phi,t_{\mu}^{a})Q_{\mu\nu}^{a}(\tilde{f},\phi,t_{\mu}^{a}) \right\rangle _{\tilde{f}} = 0.
\end{equation}
This follows from the first of equation (52). The term quartic in $ t_{\mu}^{a} $ is also zero, i.e.,
\begin{equation}\label{56}
g^{2}\left\langle Q_{\mu\nu}^{a}(\tilde{f},\phi,t_{\mu}^{a}) Q_{\mu\nu}^{a}(\tilde{f},\phi,t_{\mu}^{a}) \right\rangle _{\tilde{f}} = 0.
\end{equation}
Same reasoning as the other vanishing terms. The quadratic in $ t_{\mu}^{a} $ term given by
\begin{equation}\label{57}
\left\langle Z_{\mu\nu}^{a}(\tilde{f},\phi)Q_{\mu\nu}^{a}(\tilde{f},\phi,t_{\mu}^{a}) \right\rangle_{\tilde{f}} = 0,
\end{equation}
since $ Z(\tilde{f}) = 0 $ and all the averages  of terms with $ \phi $ are of the form given by the first of equation (52). 

	The only non-vanishing term is quadratic in $ t_{\mu}^{a} $ and is given by
\begin{equation}\label{58}
\left\langle L_{\mu\nu}^{a}(\tilde{f},\phi,t)L_{\mu\nu}^{a}(\tilde{f},\phi,t) \right\rangle_{\tilde{f}} \neq 0.
\end{equation}
Using equations (16b) and (17), we find the components of the above term.
\begin{equation}\label{59}
\left\langle R^{ab}(f)R^{ac}(f)\right\rangle _{\tilde{f}}(\partial_{\mu}t_{\nu}^{b} - \partial_{\nu}t_{\mu}^{b})(\partial_{\mu}t_{\nu}^{c} - \partial_{\nu}t_{\mu}^{c}) = \frac{1}{12\sqrt{2}\pi}(\partial_{\mu}t_{\nu}^{a} - \partial_{\nu}t_{\mu}^{a})^{2}.
\end{equation}
To arrive at this term, it is important to notice that all the terms with $ \phi $ are of the type  given by equation (52a) while the non-vanishing term is of the type given by equation (52b) with $ n = 2 $ and $ m = 3 $. There are no divergent terms.
The next non-vanishing term is
\begin{equation}\label{60}
2\left\langle R^{ab}(f)Y^{acd}(f)(\partial_{\mu}f^{c}t_{\nu}^{d} - \partial_{\nu}f^{c}t_{\mu}^{d})\right\rangle _{\tilde{f}}(\partial_{\mu}t_{\nu}^{b} - \partial_{\nu}t_{\mu}^{b}) = \frac{1}{12\sqrt{2}\pi}\frac{14}{16}\frac{\alpha}{l_{0}}(\partial_{\mu}t_{\nu}^{a} - \partial_{\nu}t_{\mu}^{a})(\frac{x_{\mu}}{x}t_{\nu}^{a} - \frac{x_{\nu}}{x}t_{\mu}^{a}).
\end{equation}     
The evaluation of equation (60) made use of equation (49). The last non-vanishing term is given by
\begin{equation}\label{61}
\left\langle Y^{abc}Y^{ade}(\partial_{\mu}f^{b}t_{\nu}^{c}-\partial_{\nu}f^{b}t_{\mu}^{c})(\partial_{\mu}f^{d}t_{\nu}^{e}-\partial_{\nu}f^{d}t_{\mu}^{e})\right\rangle _{\tilde{f}} = \frac{1}{12\sqrt{2}\pi}(1.936)(\frac{x_{\mu}}{x}t_{\nu}^{a}-\frac{x_{\nu}}{x}t_{\mu}^{a})^{2}.
\end{equation}
The evaluation of this term is a bit more involved because of equation (49). There is one term that involves $ \tilde{f}^{b}(x+\frac{\textit{l}}{\alpha})\tilde{f}^{d}(x+\frac{\textit{l}}{\alpha}) $, with an average that diverges as $ \sigma^{-1} $ as $ \sigma \rightarrow 0 $. But this divergence is cancelled by the terms involving purely $ \tilde{f}(x) $, with with average that goes to zero as $\sigma $. Then there is another term involving only $ \tilde{f}(x) $ fields and their average is finite. The combination of these two contributions result in the value given by equation (61).

	Taking into account equations (59) to (61), the effective action for the 'gluons' is given by
\begin{equation}\label{62}
\begin{split}
S_{eff}(t_{\mu}^{a})& = \frac{1}{4}\frac{1}{12\sqrt{2}\pi}\int d^{4}x \bigg\lbrace (\partial_{\mu}t_{\nu}^{a} - \partial_{\nu}t_{\mu}^{a})^{2} \\ 
&\quad + \frac{14}{16}\frac{\alpha}{\textit{l}}(\partial_{\mu}t_{\nu}^{a} - \partial_{\nu}t_{\mu}^{a})(\frac{x_{\mu}}{x}t_{\nu}^{a} -\frac{x_{\nu}}{x}t_{\mu}^{a})+ 1.936\frac{\alpha^{2}}{\textit{l}^{2}}(\frac{x_{\mu}}{x}t_{\nu}^{a} - \frac{x_{\nu}}{x}t_{\mu}^{a})^{2}\bigg\rbrace.
\end{split}
\end{equation}
This shows that the gluons are no longer self coupled but their effective Lagrangian breaks translation invariance. This is consistent with the fact that we are describing physics in the confining region, which is only valid to within the length scale and thus we cannot translate beyond this length scale. But where is the confinement of the 'gluons'? Note that if we define a new Abelian gauge field by
\begin{equation}\label{63}
t_{\mu}^{a} = \sqrt{12\sqrt{2}\pi}\exp {(-\dfrac{7 \alpha}{16 l_{0}}x)}V_{\mu}^{a}(x),
\end{equation}
where the x in the exponential is the Euclidean length of the four vector $ x_{\mu} $. It is clear from this that the 'gluon' $ t_{\mu}^{a} $, since it is exponentially decaying, must be confined to within a distance scale defined by the length scale $ \textit{l} $. The effective action for the Abelian gauge field $ V_{\mu}^{a} $ is
\begin{equation}\label{64}
S_{eff}(V_{\mu}^{a}) = \int d^{4}x \exp {(-\frac{14}{16}\frac{\alpha}{\textit{l}}x)} \left\lbrace \frac{1}{4}(\partial_{\mu}V_{\nu}^{a} - \partial_{\nu}V_{\mu}^{a})^{2} + \frac{1}{2}m^{2}V_{\mu}^{a}V_{\mu}^{a} \right\rbrace,
\end{equation}
where the Abelian field $ V_{\mu}^{a} $ has mass m equal to $ \sqrt{1.745}\frac{\alpha}{\textit{l}} $. We also made use of $ x \cdot V^{a} = 0 $, the radial gauge, which follows from equation (13b) with $ f^{a} = \tilde{f}^{a} $ and equation (49b). 

	The effective action suggests that the Abelian fields are in a background metric but they are not because the derivatives are ordinary and not covariant derivatives. The action breaks translation symmetry because of the exponential term. The field equation for the Abelian fields is
\begin{equation}\label{65}
\partial_{\mu}(\partial_{\mu}V_{\nu}^{a} - \partial_{\nu}V_{\mu}^{a}) + m^{2}V_{\nu}^{a} - \frac{x_{\mu}}{x}(\partial_{\mu}V_{\nu}^{a} - \partial_{\nu}V_{\mu}^{a}) = 0.
\end{equation}   	         
Because of the last term in equation (65), the massive Abelian fields $ V_{\mu} $ are no longer transverse. Equation (65) is to be solved subject to the boundary conditions that (1) as $ x \rightarrow 0 $, the Abelian fields must go like the transverse gauge potentials in perturbative QCD and (2) as $ x \rightarrow \textit{l} $, the Abelian fields must vanish. These two conditions must be enough to determine the allowed masses of the 'Abelian gluon' fields. In short we will be able to solve for the mass gap. But solving equation (65) in 4D is rather involved so we will solve it in 2D to show how the mass gap arises.

	In 2D and using plane polar coordinates, there is only one degree of freedom, the $ V_{\theta} $ and if we assume it is purely a function of the radial coordinate r, we find that it satisfies the second order, linear, ordinary differential equation with non-constant coefficients
\begin{equation}\label{66}
\dfrac{d^{2}V_{\theta}}{dr^{2}} + (\frac{1}{r}-a)\dfrac{dV_{\theta}}{dr} + (m^{2}-\frac{a}{r}-\frac{1}{r^{2}})V_{\theta} = 0.
\end{equation}
It is important to keep in mind that equation (66) is valid in the confinement regime, i.e., $ [r_{asy},\textit{l}] $ where $ r_{asy} $ is the largest asymptotic freedom distance scale and $ \textit{l} $ is the confinement length. Below this regime is asymptotic freedom  regime where the linear part of the non-linear gauge is valid while above is the hadronization distance scale of pion-nucleon interaction.
 
	We can simplify the equation by defining $ V_{\theta}(r) = \frac{1}{r}F(r) $ and $ F(r) $ satisfies
\begin{equation}\label{67}
\dfrac{d^{2}F}{dr^{2}} - (a+\frac{1}{r})\dfrac{dF}{dr} + m^{2}F = 0.
\end{equation}
The solution to equation (67) is discussed in Appendix B and is given by
\begin{subequations}\label{68}
\begin{gather}
F(r) = f(r) \exp {(\frac{ar}{2})}\sin {\beta r},\label{first}\\
f(r) = \sum_{0}^{\infty} a_{i}(\beta r)^{i},
\end{gather}
\end{subequations}
where $ \beta = \sqrt{m^{2} - \frac{a^{2}}{4}} $ and the expansion coefficients are $ a_{0} = 0 $, $ a_{1} $ is undetermined and all $ a_{j} $ for $ j \geqslant 2 $ are given in terms of $ a_{1} $ with declining values. For example, $ a_{2} = \frac{1}{3}\kappa a_{1} $, $ a_{3} = \frac{1}{24}(1 + \kappa^{2})a_{1} $ and $ a_{4} = (\frac{11}{1080}\kappa + \frac{1}{360}\kappa^{3})a_{1} $, where $ \kappa = \frac{a}{2\beta} = $. At the confinement distance $ \textit{l} $, $ F(\textit{l}) $ is equal to zero, giving $ \beta \textit{l} = n\pi $, with $ n = 1,2,.... $. However, as Appendix B will show, we can only have $ n = 1 $ resulting in the parameter $ \alpha $, defined in the definition of the derivative for the white noise given in equation (49b) with value given by 
\begin{equation}\label{69}
\alpha = \dfrac{\pi}{\sqrt{1.745 - \frac{1}{4}(\frac{14}{16})^{2}}}.
\end{equation}

	The mass of the 'Abelian gluon' $ V_{\mu}^{a} $ is $ m^{2} = 1.745 \frac{\alpha^{2}}{\textit{l}^{2}} $. Thus, in $ QCD_{2} $, there is only one massive state inside the confinement regime. Thus, we have exhibited the mass gap in $ QCD_{2} $. It is more difficult to exhibit the mass gap in $ QCD_{4} $ because equation (65) is more formidable there. 
	Before we continue with the addition of quarks, it is important to point out that the 'gluon' condensate, which is given by
\begin{equation}\label{70}
\left\langle F^{a}_{\mu\nu}F^{a}_{\mu\nu}\right\rangle _{\tilde{f}} = \exp {(-\frac{14}{16}\frac{\alpha}{l_{0}}x)} \left\lbrace \frac{1}{4}(\partial_{\mu}V_{\nu}^{a} - \partial_{\nu}V_{\mu}^{a})^{2} + \frac{1}{2}m^{2}V_{\mu}^{a}V_{\mu}^{a} \right\rbrace_{\tilde{V}}, 
\end{equation}
where $ \tilde{V} $ is a solution to equation (65). Unfortunately, we do not have the solution to equation (65) in 4D, but we do have the solution in 2D as given in Appendix B. The 2D gluon condensate is given by 
\begin{equation}\label{71}
\left\langle F^{2}(0) \right\rangle _{vac} = a_{1}^{2}\beta^{3},
\end{equation}
which is clearly not zero, thus consistent with SVZ \cite{Shifman}.

	To end this section, we restate  that we have shown gluon confinement. We have also computed, in $ QCD_{2} $ the mass gap and show explicitly that the gluon vacuum condensate is not zero, consistent with SVZ. Carrying this out in 4D is the next step, albeit a more challenging computation.  

\section{\label{sec:level7}Dynamical Quark Confinement} 

	Finally, we prove that the quarks that constitute mesons and baryons, not just the static, massive quarks considered in the Wilson loop, are confined. We begin with the path integral with quarks added and it is given by
\begin{equation}\label{72}
\begin{split}
W& = \int (dt_{\mu}^{a})(df^{a})(d\Psi)(d\bar{\Psi})\delta(\partial \cdot t^{a}-\frac{1}{g\textit{l}^{2}}f^{a})\delta(\rho^{a})\\
&\quad det^{-4}(1 + \vec{f}\cdot\vec{f})det\Theta \exp {[-(S_{YM} + S_{fermion})]}.
\end{split}
\end{equation}
The first thing we do is rewrite the delta functionals as
\begin{subequations}\label{73}
\begin{gather}
\delta(\partial \cdot t^{a}-\frac{1}{g\textit{l}^{2}}f^{a})=det\left[ \dfrac{1}{(1+\vec{f}\cdot\vec{f})^{j}}\right] \delta((\dfrac{1}{(1+\vec{f}\cdot\vec{f})^{j}})(\partial \cdot t^{a}-\frac{1}{g\textit{l}^{2}}f^{a})),\label{first}\\
\delta(\rho^{a})=det\left[ \dfrac{1}{(1+\vec{f}\cdot\vec{f})^{k}}\right]\delta(\dfrac{1}{(1+\vec{f}\cdot\vec{f})^{k}}\rho^{a}),
\end{gather}
\end{subequations}
where j and k are integers that can be conveniently chosen to make certain stochastic averages vanish. The path-integral (72) can be written as
\begin{equation}\label{74}
W = \int (dt_{\mu}^{a})(df^{a})(d\Psi)(d\bar{\Psi})(du^{a})(d\bar{u}^{a}) \exp {(-S)},
\end{equation}
where $ S' $ is given by
\begin{subequations}\label{75}
\begin{gather}
S = S_{YM} + S_{fermion} + S_{gf} + S_{ghosts},\label{first}\\
S_{gf} = \int d^{4}x \left\lbrace  \frac{1}{\alpha}\dfrac{1}{(1+\vec{f}\cdot\vec{f})^{2j}}(\partial \cdot t^{a}-\frac{1}{g\textit{l}^{2}}f^{a})^{2} + \frac{1}{\beta}\dfrac{1}{(1+\vec{f}\cdot\vec{f})^{2k}}(\rho^{a})^{2}\right\rbrace,\label{second}\\
S_{ghosts} = \int d^{4}x \dfrac{1}{(1+\vec{f}\cdot\vec{f})^{j+k+4}} \bar{u}^{a}\Theta^{ab}u^{b}.
\end{gather}
\end{subequations}
In equation (75b), we need to take both $ \alpha $ and $ \beta  \longrightarrow 0 $. The integer parameters j and k are chosen so that the stochastic averages involving $ S_{gf} $ and $ S_{ghosts} $ will lead to integrals that will satisfy the first of equation (52), i.e., vanishing stochastic averages. Thus, we only need to look at stochastic averages involving $ S_{YM} $ and $ S_{fermion} $.

	Now we take the stochastic average of the path-integral (74), which we carry out by using equation (47) and the path-integral in $ f^{a} $ becomes a path-integral in $ \phi^{a} $ and the stochastic average in $ \tilde{f}^{a} $ makes use of the distribution given by equation (23). Expanding the exponential of $ S $, we need to evaluate the stochastic averages of powers of $ S $. These are 
\begin{subequations}\label{76}
\begin{gather}
\left\langle S_{ghosts} \right\rangle_{\tilde{f}} = 0,\label{first}\\
\left\langle S_{gf} \right\rangle_{\tilde{f}}  = 0,\label{second}\\
\left\langle S_{YM} \right\rangle_{\tilde{f}} = S_{eff}(V_{\mu}^{a}),\label{third}\\
\left\langle S_{fermion} \right\rangle_{\tilde{f}} = \int d^{4}x \bar{\Psi}i\gamma_{\mu}\left( \partial_{\mu} -gT^{a}\left\langle A^{a}_{\mu}(x) \right\rangle _{\tilde{f}}\right) \Psi.
\end{gather}
\end{subequations}
The right hand side of equation (76c) is given by equation (64). We evaluate equation (76d) by making use of equation (12), the background decomposition given by (47) and equation (52) resulting in
\begin{equation}\label{77}
\left\langle A_{\mu}^{a}(x) \right\rangle_{\tilde{f}} = -\frac{1}{12\pi}(\partial_{\mu}\phi^{a} + t_{\mu}^{a}) + \frac{1}{3\pi}\frac{\alpha}{\textit{l}}\frac{x_{\mu}}{x}\phi.
\end{equation}
We can make use of equation (63) to express (77) in terms of the Abelian field $ V_{\mu}^{a} $ and it will confine the quark-'gluon' interaction to the confinement length $ \textit{l} $. Putting things together, we find
\begin{equation}\label{78}
\begin{split}
\left\langle S \right\rangle_{\tilde{f}}& = \int d^{4}x \bigg\lbrace \exp {(-\frac{14}{16}\frac{\alpha}{l_{0}}x)} \left[\frac{1}{4}(\partial_{\mu}V_{\nu}^{a} - \partial_{\nu}V_{\mu}^{a})^{2} + \frac{1}{2}m^{2}V_{\mu}^{a}V_{\mu}^{a}\right] \\
& \quad + \bar{\Psi}i\gamma_{\mu}\partial_{\mu}\Psi + ig\frac{1}{\sqrt{12\sqrt{2}\pi}}\exp {(-\dfrac{7 \alpha}{16 l_{0}}x)}V_{\mu}^{a}(x)\bar{\Psi}(x)T^{a}\gamma_{\mu}\Psi(x)\\
& \quad -ig\frac{1}{3\pi} \phi^{a}\left[ \partial_{\mu}\left( \bar{\Psi}\gamma_{\mu}T^{a}\Psi\right) + 4\frac{\alpha}{\textit{l}}\frac{x_{\mu}}{x}\bar{\Psi}\gamma_{\mu}T^{a}\Psi \right]\bigg\rbrace 
\end{split}
\end{equation}

	Now we go to the stochastic averages of the higher order terms in $ \exp {(-S)} $. In previous papers \cite{Magpantay6}, \cite{Magpantay7}, the focus was on $ \left\langle S^{2} \right\rangle_{\tilde{f}} $ and in particular only on $ \left\langle A_{\mu}^{a}(x)A_{\nu}^{b}(y) \right\rangle_{\tilde{f}} $. We made a short-cut in deriving the answer by making use of equation (20) and (21), which gave the result of fermions with a non-local four-fermi interaction with linear potential. This and equation (78) clearly show the confinement of dynamical quarks. In the following, we clearly show these two results by evaluating the higher order terms in the stochastic averaging. 

	We are computing the stochastic average of $ S^{2} $ and it is given by
\begin{equation}\label{79}
\left\langle S^{2} \right\rangle_{\tilde{f}} = \int \int d^{4}x d^{4}y \left\langle \textit{L}(\Phi(x),\tilde{f}(x))\textit{L}(\Phi(y),\tilde{f}(y))\right\rangle_{\tilde{f}},
\end{equation}
where $ \textit{L}(x) = \textit{L}(\Phi(x),\tilde{f}(x)) = \textit{L}_{YM}(t_{\mu}^{a},\phi^{a},\tilde{f}^{a}) + \textit{L}_{fermion} + \textit{L}_{ghosts} + \textit{L}_{gf} $, and the Lagrangian terms are given by the actions in equation (75). Collectively, we represent all the fields, other than the stochastic background $ \tilde{f}^{a} $, by the field $ \Phi $. Now we make use of the following result in noise averaging given in \cite{Hanggi}
\begin{equation}\label{80}
\begin{split}
\left\langle \textit{L}(x)\textit{L}(y) \right\rangle_{\tilde{f}}& = \left\langle \textit{L}(x) \right\rangle_{\tilde{f}} \left\langle \textit{L}(y) \right\rangle_{\tilde{f}} + \int_{x}^{y} dx_{1} \int_{x}^{y} dy_{1} \left\langle \dfrac{\delta \textit{L}(\Phi(x),\tilde{f}(x))}{\delta \tilde{f}^{a}(x_{1})} \right\rangle_{\tilde{f}} \left\langle \dfrac{\delta \textit{L}(\Phi(y),\tilde{f}(y))}{\delta \tilde{f}^{b}(y_{1})} \right\rangle_{\tilde{f}} \left\langle \tilde{f}^{a}(x_{1})\tilde{f}^{b}(y_{1})\right\rangle_{\tilde{f}} \\
& \quad + \frac{1}{2} \int dx_{1} \int dx_{2} \int dy_{1} \int dy_{2} \left\langle \dfrac{\delta^{2} \textit{L}(\Phi(x),\tilde{f}(x))}{\delta \tilde{f}^{a}(x_{1})\delta \tilde{f}^{b}(x_{2})} \right\rangle_{\tilde{f}} \left\langle \dfrac{\delta^{2} \textit{L}(\Phi(y),\tilde{f}(y))}{\delta \tilde{f}^{c}(y_{1})\delta \tilde{f}^{d}(y_{2})} \right\rangle_{\tilde{f}} \\
& \quad \cdot \left\lbrace \left\langle \tilde{f}^{a}(x_{1})\tilde{f}^{c}(y_{1})\right\rangle_{\tilde{f}} \left\langle \tilde{f}^{b}(x_{2})\tilde{f}^{d}(y_{2})\right\rangle_{\tilde{f}} + \left\langle \tilde{f}^{a}(x_{1})\tilde{f}^{d}(y_{2})\right\rangle_{\tilde{f}} \left\langle \tilde{f}^{b}(x_{2})\tilde{f}^{c}(y_{1})\right\rangle_{\tilde{f}}\right\rbrace \\   
& \quad + .....
\end{split}
\end{equation}
The first term of the RHS of equation (80) is just the square of the stochastic average given by equation (76) and it is part of the quadratic term in $ \exp {(-S_{eff}(V_{\mu}^{a},\tilde{\psi},\psi))} $. The second term of equation (80) will give the non-local four-fermi with linear potential interaction and is part of the linear term in the expansion of $ \exp {(-S_{eff}(V_{\mu}^{a},\tilde{\psi},\psi))} $ along with equation (76). Starting from the third term onwards, given by the ellipsis, are all zeros as can easily seen.

	Taking the second derivative of all the components of  $ \textit{L} $ with respect to $ \tilde{f} $ will result in stochastic averages that follow the first of equation (52), thus zero. This results in zero value for the third and succeeding terms of equation (80). Let us write down the detailed expression of the second term of (80) as part of the stochastic average given in (79).
\begin{equation}\label{81}
\begin{split}
\int d^{4}x \int d^{4}y \left\lbrace (80)_{2^{nd}} \right\rbrace & = \int d^{4}x \int d^{4}y  (\bar{\Psi}\gamma_{\mu}T^{a}\Psi)_{x}\left\langle \dfrac{\delta A_{\mu}^{a}(x)}{\delta \tilde{f}^{c}(x)} \right\rangle_{\tilde{f}} \\
& \quad \cdot \left\lbrace \int_{x}^{y} \int_{x}^{y} dx_{1} dy_{1} \left\langle \tilde{f}^{c}(x_{1})\tilde{f}^{d}(y_{1}) \right\rangle_{\tilde{f}}\right\rbrace \\
& \quad \cdot \left\langle \dfrac{\delta A_{\nu}^{b}(y)}{\delta \tilde{f}^{d}(y)} \right\rangle_{\tilde{f}} (\bar{\Psi}\gamma_{\nu}T^{b}\Psi)_{y}.
\end{split}
\end{equation}
Equation (81) follows from the fact that the terms in $ \textit{L}_{YM} $ that have non-zero stochastic average given by equations (59), (60) and (61), i.e., their stochastic averages follow the second of equation (52), once differentiated by $ \tilde{f} $ will now have stochastic averages that follow the first of (52). The vanishing of the terms that come from differentiating by $ \tilde{f} $ the terms from $ \textit{L}_{ghosts} $ and $ \textit{L}_{gf} $ are trivial because $ \textit{L}_{ghosts} $ and $ \textit{L}_{gf} $ have zero stochastic averages themselves.

	Notice that we have shifted from differentiating with respect to $ x_{1}, y_{1} $ in the functional derivatives in equation (80) to x or y in equation (81). The reason for this is when $ x_{1} \neq x $ or $ y_{1} \neq y $, the functional derivatives in equation (80) are zero. In ordinary functions, the functional derivative $ \frac{\delta f(x)}{\delta f(x_{1}} = \delta (x-x_{1}) $. But $ \tilde{f}(x) $ is a white noise, which makes it nowhere differentiable, making $ \delta \tilde{f}(x) $ ill-defined and the functional derivative $ \frac{\delta \tilde{f}(x)}{\delta \tilde{f}(x_{1}} $ more ill-defined. In the literature \cite{Hanggi}, this derivative is defined in such a way that $ \frac{\delta \tilde{f}(x)}{\delta \tilde{f}(x_{1}} = \delta (x-x_{1}) $ but this seems to be questionable as discussed above. This is the reason why the differentiation in equation (80) was done in x or y leading to equation (81). 
	
	  The stochastic average of $ A_{\mu}^{a}(x) $ is given by equation (77) and one would expect that the stochastic average of the derivative of this term with respect to $ \tilde{f} $ will be zero. However it is not as can be seen from the fact that 
\begin{equation}\label{82}
\begin{split}
A_{\mu}^{a}(\tilde{f})& = \dfrac{1}{(1+\tilde{\vec{f}}\cdot\tilde{\vec{f}})_{x}}\left( \delta^{ab}+\epsilon^{abc}\tilde{f}^{c}+\tilde{f}^{a}\tilde{f}^{b}\right)_{x} \left[\frac{1}{g}\frac{x_{\mu}}{x}\frac{\alpha}{\textit{l}} \left( \tilde{f}^{b}(x+\frac{\textit{l}}{\alpha})-\tilde{f}^{b}(x) \right) \right]\\
& = (\frac{1}{g}\frac{x_{\mu}}{x}\frac{\alpha}{\textit{l}})\left[ \dfrac{1}{(1+\tilde{\vec{f}}\cdot\tilde{\vec{f}})}_{x} \left( \delta^{ab}+\epsilon^{abc}\tilde{f}^{c}+\tilde{f}^{a}\tilde{f}^{b} \right)_{x} \tilde{f}^{b}(x+\frac{\textit{l}}{\alpha}) - \tilde{f}^{a}(x)\right], 
\end{split}
\end{equation}
where we made use of equation (49). The second of equation (82) clearly shows that $ \left\langle A_{\mu}^{a}(\tilde{f})\right\rangle_{\tilde{f}} = 0 $ not because of the first of equation (52) but because $ \left\langle \tilde{f}^{a}(x)\right\rangle_{\tilde{f}} = 0 $. Also from the second of equation (82), we get
\begin{equation}\label{83}
\left\langle \dfrac{\delta A_{\mu}^{a}(\tilde{f}(x))}{\delta \tilde{f}^{c}(x)}\right\rangle_{\tilde{f}} = \delta^{ac}(\frac{1}{g}\frac{x_{\mu}}{x}\frac{\alpha}{\textit{l}}).
\end{equation}

	To complete equation (81), we need to evaluate its middle component. We can do this two ways. One is by noting that 
\begin{equation}\label{84}
\left\langle \tilde{f}^{c}(x_{1})\tilde{f}^{d}(y_{1}) \right\rangle_{\tilde{f}} = \delta^{cd}\delta(x_{1}-y_{1}),
\end{equation}
and upon integrating $ x_{1} $ and $ y_{1} $ yields the linear potential. Or we can integrate first the white noise to yield a Wiener process, i.e., 
\begin{equation}\label{85}
\int_{x}^{y} \tilde{f}^{c}(x_{1}) dx_{1} = W^{c}(y) - W^{c}(x),
\end{equation}
and make use of 
\begin{equation}\label{86}
\left\langle W^{c}(x)W^{d}(y) \right\rangle = \delta^{cd} min(x,y).
\end{equation}
The result is the same,we get the linear potential.

	Taking all these into account,  we can define then the action for the non-local four-fermi interaction term with linear potential
\begin{equation}\label{87}
\begin{split}
S_{NLFF}& = \int d^{4}x \int d^{4}y \left\lbrace (80)_{2^{nd}} \right\rbrace \\ 
& \quad = \int d^{4}x \int d^{4}y  (\bar{\Psi}\gamma_{\mu}T^{a}\Psi)_{x}(\frac{1}{g}\frac{\alpha}{\textit{l}})^{2} \eta_{\mu}(x) \eta_{\nu}(y) \vert y - x \vert (\bar{\Psi}\gamma_{\nu}T^{a}\Psi)_{y},
\end{split}
\end{equation}
where $ \eta_{\mu}(x) = \frac{x_{\mu}}{x} $ are the directional unit vectors in spherical coordinates in $ R^{4} $. Equation (81) gives the non-local four-fermi term with linear potential, which was derived in \cite{Magpantay7} by using equation (20). Going back to equation {80). it can now be written as 
\begin{equation}\label{88}
\left\langle S^{2} \right\rangle_{\tilde{f}} = \left\langle S \right\rangle_{\tilde{f}}^{2} + S_{NLFF}.
\end{equation}

	The next stochastic average is that of the cubic term, i.e., $ \left\langle S^{3} \right\rangle_{\tilde{f}} $. Following equation (80) and the fact that the stochastic average with two functional derivatives of any $ \textit{L} $ terms are all zeroes, we find
\begin{equation}\label{89}
\left\langle S^{3} \right\rangle_{\tilde{f}} = \left\langle S \right\rangle_{\tilde{f}}^{3} + 3 \left\langle S \right\rangle_{\tilde{f}} S_{NLFF}.
\end{equation}

	Now we add the contributions of the stochastic averages of each power and we find that
\begin{subequations}\label{90}
\begin{gather}
\left\langle \exp {(-S)} \right\rangle_{\tilde{f}} = \exp {(-S_{eff}(V,\Psi,\phi))},\label{first}\\
S_{eff}(V,\Psi,\phi) = \left\langle S \right\rangle_{\tilde{f}} + S_{NLFF},
\end{gather}
\end{subequations}
where $ \left\langle S \right\rangle_{\tilde{f}}
$ is given by equation (78). Finally, we can integrate out the scalars $ \phi^{a} $ and we get the path-integral for the fermions and Abelian 'gluons' given by
\begin{equation}\label{91}
W = \int (dV_{\mu}^{a})(d\bar{\Psi})(d\Psi) \delta\left( \partial_{\mu}\left( \bar{\Psi}\gamma_{\mu}T^{a}\Psi\right) + 4\frac{\alpha}{\textit{l}}\frac{x_{\mu}}{x}\bar{\Psi}\gamma_{\mu}T^{a}\Psi\right)	 \exp {(-\textbf{S}(V,\bar{\Psi},\Psi))},
\end{equation}
where 
\begin{equation}\label{92}
\begin{split}
\textbf{S}(V,\bar{\Psi},\Psi))& = \int d^{4}x \bigg\lbrace \exp {(-\frac{14}{16}\frac{\alpha}{l_{0}}x)} 
\left[\frac{1}{4}(\partial_{\mu}V_{\nu}^{a} - \partial_{\nu}V_{\mu}^{a})^{2} + \frac{1}{2}m^{2}V_{\mu}^{a}V_{\mu}^{a}\right] \\
& \quad + \bar{\Psi}i\gamma_{\mu}\partial_{\mu}\Psi + ig\frac{1}{\sqrt{12\sqrt{2}\pi}}\exp {(-\dfrac{7 \alpha}{16 l_{0}}x)}V_{\mu}^{a}(x)\bar{\Psi}(x)\gamma_{\mu}\Psi(x)\bigg\rbrace  \\
& \quad + \int d^{4}x \int d^{4}y  (\bar{\Psi}\gamma_{\mu}T^{a}\Psi)_{x}(\frac{1}{g}\frac{\alpha}{\textit{l}})^{2} \eta_{\mu}(x) \eta_{\nu}(y) \vert y - x \vert (\bar{\Psi}\gamma_{\nu}T^{a}\Psi)_{y} . 
\end{split}
\end{equation}

	Equations (91) and (92) describes the physics of quarks and 'gluons' in the confining region. Note, the original vector field $ A_{\mu}^{a} $ does not appear and is replaced by the Abelian gluons $ V_{\mu}^{a} $, which is well confined in the confining region (action is exponentially damped) and exhibits a mass gap. In $ QCD_{2} $, we explicitly showed in Section VI that there is only one massive state with mass given in terms of the confining length ( see equation (69) and the discussion immediately after that). For $ QCD_{4} $, the mass gap is rather involved and still needs to be worked at. 
	
	The quarks are also clearly confined. The local quark-Abelian gluon interaction is also exponentially damped. The quark currents are confined within the confinement regime as can be seen from the delta function, whose solution is exponentially damped. Lastly, the quarks experience a non-local four Fermi interaction with a linear potential. In the limit of heavy quarks, this will give the phenomenology of quarks in a linear potential, the same as the Wilson loop results.
            
\begin{acknowledgments}

I am grateful to Felicia G. Magpantay for discussions regarding Appendix B and and cleaning my latex file. I would like to thank Mike Solis and Eric Galapon for help in including appendices and figures in latex, and Perry Esguerra for a reference source.

\end{acknowledgments}


\renewcommand{\theequation}{A-\arabic{equation}}
\setcounter{equation}{0}  
\setcounter{section}{0}
\section*{Appendix A. Non-Singular Nature of $\Theta$}
\setcounter{section}{0}

	The operator $ \Theta^{ab} $ defined by equation (8b) arose when we asked if there is a transformation that will connect the boundaries $ \textit{l}_{f} $ and $ \textit{l}_{f'} $. We already know from the discussions that fields on $ \textit{l}_{f} $ is one orbit and by similar arguments the same with fields on  $ \textit{l}_{f'} $. The solution to equation (8a) will answer if these two orbits are linked by a gauge transformation, i.e., the two orbits actually comprise one orbit. And we find that there is no gauge parameter $ \Lambda $ that will take us from $ \textit{l}_{f} $ to $ \textit{l}_{f'} $. And the proof is precisely because the operator $ \Theta^{ab} $ is non-singular so equation (8a) does not have a solution.
	
	The proof is straightforward. The first term of $ \Theta $ is a Laplacian squared, thus it has four spatial derivatives. This operator has a zero mode which is precisely $ \partial \cdot A $. The second term of $ \Theta $ is first-order in spatial derivatives and thus can be considered a small correction. First-order perturbation theory says the zero mode of $ \Theta $ must be 
\begin{equation}\label{A.1}
\Lambda^{a} = \partial \cdot A^{a} + z^{a}, 
\end{equation}
where $ z^{a} $ is solved from
\begin{equation}\label{A.2}
(D \cdot \partial)^{ab}(\partial \cdot D)^{bc}z^{c} = g\epsilon^{abc}\partial(\partial \cdot A^{b})\cdot(D^{cd})(\partial \cdot A^{d}).
\end{equation}
For this equation to have a non-trivial solution for $ z^{a} $, the zero mode of the operator at the LHS of equation (A.2), which is $ \partial \cdot A $ must be orthogonal to the source, the RHS of equation (A.2). This means
\begin{equation}\label{A.3}
\int d^{4}x (\partial \cdot A^{a})\epsilon^{abc}\partial(\partial \cdot A^{b})\cdot(D^{cd})(\partial \cdot A^{d}) = 0,
\end{equation}
which clearly cannot be satisfied. Thus, $ \Theta $ does not have a zero mode and the operator is non-singular. It also means the orbits $ \textit{l}_{f} $ and $ \textit{l}_{f'} $ are separate and distinct. 

\renewcommand{\theequation}{B-\arabic{equation}}
\setcounter{equation}{0}  
\setcounter{section}{0}

\section*{Appendix B. Mass Gap in QCD$_{2}$}
\setcounter{section}{0}

	In 2D pure QCD, we have to solve equation (66). Defining $ V_{\theta}(r) = \frac{1}{r}F(r) $, we find equation (67). Without the $ \frac{1}{r} $ term, this equation is solved by 
\begin{equation}\label{B.1}
F_{0}(r) = \exp {(\frac{a}{2}r)}\sin (\beta r),
\end{equation}
where $ \beta = \sqrt{m^{2}-\frac{a^{2}}{4}} $. I then propose that equation (68) is solved by
\begin{equation}\label{B.2}
F(r) = f(r)F_{0}(r),
\end{equation}
resulting in the following equation for $ f(r) $
\begin{equation}\label{B.3}
\dfrac{d^{2}f}{dr^{2}} + \left[ 2 \dfrac{d\ln F_{0}}{dr} - (a+\frac{1}{r})\right] \dfrac{df}{dr} -\frac{1}{r}\dfrac{d\ln F_{0}}{dr} = 0.
\end{equation}. 
To solve this equation, we make use of the following expansion, see  page 46 of \cite{Gradshteyn}
\begin{equation}\label{B.4}
\ln sin(\beta r) = \ln (\beta r) + \sum_{k=1}^{\infty}	\dfrac{(-1)^{k}2^{2k-1}B_{2k}}{k(2k)!}(\beta r)^{2k},
\end{equation}
where $ B_{2k} $ are Bernoulli numbers and the series expansion is valid for $ \beta r < \pi $. From this we get
\begin{equation}\label{B.5}
\dfrac{d\ln F_{0}}{dr} = \frac{a}{2}+\frac{1}{r}-\dfrac{2\beta^{2}r}{6}-\dfrac{4\beta^{4}r^{3}}{180}-\dfrac{6\beta^{6}r^{5}}{2835}-.....,
\end{equation}	
where we made use f the first few Bernoulli numbers. 
Substitute this in (B.3), we get the following differential equation
\begin{equation}\label{B.6}
\begin{split}
0& = \dfrac{d^{2}f}{dr^{2}} + \left[ \frac{\beta}{\beta r}-\frac{4}{6}\beta (\beta r)-\frac{8}{180}\beta(\beta r)^{3}-\frac{12}{2835}\beta(\beta r)^{5}-...\right]\dfrac{df}{dr}\\ 
& \quad -\left[ \frac{\beta^{2}}{\beta r}^{2}+\frac{a\beta}{2}\frac{1}{\beta r}-\frac{2\beta^{2}}{6}-\frac{4\beta^{2}}{180}(\beta r)^{2}-\frac{6\beta^{2}}{2835}(\beta r)^{4}-....\right]f.
\end{split}
\end{equation}
Equation (B.5) suggests the series solution given by equation (68b). The resulting recursion relations give the values of the expansion coefficients listed in the discussions after equation (68).

	This completes the solution for pure QCD2.  


\bibliographystyle{abbrvnat}
\bibliography{biblio}

\begin{thebibliography}{20}
\providecommand{\natexlab}[1]{#1}
\providecommand{\url}[1]{\texttt{#1}}
\expandafter\ifx\csname urlstyle\endcsname\relax
  \providecommand{\doi}[1]{doi: #1}\else
  \providecommand{\doi}{doi: \begingroup \urlstyle{rm}\Url}\fi

\bibitem[Amiri(2010)]{Amiri}
F.~M. B.~M. Amiri, A.
\newblock Generalization of some determinantal identities for non-square
  matrices based on radic's definition.
\newblock \emph{J. Pure Appl. Math.}, 1\penalty0 (63), 2010.

\bibitem[Dell'Antonio(1991)]{DellAntonio}
Z.~D. Dell'Antonio, G.
\newblock Every gauge orbit passes inside the gribov horizon.
\newblock \emph{Comm. Math. Physics}, 138\penalty0 (291), 1991.

\bibitem[Fritsch(1973)]{Fritsch}
G.~M. L.~H. Fritsch, H.
\newblock Advantages of the color octet gluon picture.
\newblock \emph{Phys. Lett.}, 47B\penalty0 (365), 1973.

\bibitem[Gradshteyn(1965)]{Gradshteyn}
R.~I. Gradshteyn, I.
\newblock \emph{Table of Integrals. Series and Products}.
\newblock Academic Press, NY, 1965.

\bibitem[Gribov(1978)]{Gribov}
V.~Gribov.
\newblock Quantization of non-abelian gauge theories.
\newblock \emph{Nucl. Physics B}, 139\penalty0 (1), 1978.

\bibitem[Gross(1973)]{Gross}
W.~F. Gross, D.
\newblock Ultraviolet behavior of non-abelian gauge theories.
\newblock \emph{Phys. Rev. Lett.}, 30\penalty0 (1343), 1973.

\bibitem[Hanggi(1989)]{Hanggi}
J.~P. Hanggi, P.
\newblock \emph{Colored Noise in Dynamical Systems: A Functional Approach, in
  Noise in Non-Linear Dynamical Systems}, volume~1.
\newblock Cambridge University Press, 1989.

\bibitem[Macarewicz(2014)]{Makarewicz}
P.~P. S.~D. Macarewicz, A.
\newblock Properties of the determinant of a rectangular matrix.
\newblock \emph{Annales Universitatis Marie Curie - Sklodowska, Lublin-Polonia,
  LXVIII}, 1\penalty0 (31), 2014.

\bibitem[Magpantay(1994)]{Magpantay2}
J.~Magpantay.
\newblock The coulomb gauge revisited.
\newblock \emph{Prog. of Theor. Physics}, 91\penalty0 (573), 1994.

\bibitem[Magpantay(1999{\natexlab{a}})]{Magpantay1}
J.~Magpantay.
\newblock Yang-mills theory in a non-linear gauge.
\newblock In M.~N. K. W.~K. Bernido~C., Carpio-Bernido, editor, \emph{2nd Jagna
  International Workshop on Mathematical Methods of Quantum Physics}. Gordon
  and Breach, 1999{\natexlab{a}}.

\bibitem[Magpantay(1999{\natexlab{b}})]{Magpantay3}
J.~Magpantay.
\newblock The confinement mechanism in yang-mills theory.
\newblock \emph{Modern Physics Letters A}, 14\penalty0 (447),
  1999{\natexlab{b}}.

\bibitem[Magpantay(2000)]{Magpantay4}
J.~Magpantay.
\newblock The parisi-sourlas mechanism in yang-mills theory.
\newblock \emph{International Journal of Modern Physics A}, 15\penalty0 (1613),
  2000.

\bibitem[Magpantay(2002)]{Magpantay5}
J.~A. Magpantay.
\newblock Effective "gluon" dynamics in a stochastic vacuum.
\newblock 2002.

\bibitem[Magpantay(2004{\natexlab{a}})]{Magpantay6}
J.~A. Magpantay.
\newblock A confining non-local four-fermi interaction from yang-mills theory
  in a stochastic background.
\newblock 2004{\natexlab{a}}.

\bibitem[Magpantay(2004{\natexlab{b}})]{Magpantay7}
J.~A. Magpantay.
\newblock Effective quantum dynamics of quarks and gluons in a stochastic
  background.
\newblock 2004{\natexlab{b}}.

\bibitem[McNeile(2003)]{McNeile}
C.~McNeile.
\newblock Meson and baryon spectroscopy on a lattice.
\newblock 2003.
\newblock \doi{10.1142/9789812701381_0001}.

\bibitem[Parisi(1979)]{Parisi}
S.-N. Parisi, G.
\newblock Random magnetic fields and dimensional reduction in spin systems.
\newblock \emph{Phys. Rev. Lett.}, 43\penalty0 (744), 1979.

\bibitem[Politzer(1973)]{Politzer}
H.~Politzer.
\newblock Reliable perturbative results for strong interactions.
\newblock \emph{Phys. Rev. Lett.}, 30\penalty0 (1346), 1973.

\bibitem[Radic(1979)]{Radic}
M.~Radic.
\newblock A definition of the determinant of a rectangular matrix.
\newblock \emph{Glasnik Matemaicki}, 1\penalty0 (17), 1979.

\bibitem[Shifman(1979)]{Shifman}
V.-A. Z.~V. Shifman, M.
\newblock Qcd and resonance physics.
\newblock \emph{Nucl. Physics B}, 147\penalty0 (385), 1979.

\end{thebibliography}

\end{document}